\documentclass[preprint]{elsarticle}
\biboptions{numbers,sort&compress}

\usepackage{amssymb}
\usepackage{amsmath}
\usepackage{float}
\usepackage{enumitem}
\usepackage{booktabs}
\usepackage{url}
\usepackage{hyperref}

\usepackage[table]{xcolor}
\usepackage{soul}
\sethlcolor{yellow}
\soulregister{\cite}{7}
\soulregister{\ref}{7}
\newif\ifshowrevisions
\showrevisionstrue % \showrevisionsfalse gives the clean version

\journal{Biomedical Signal Processing and Control}

\begin{document}

\begin{frontmatter}

\title{orGAN: Bridging Synthetic and Real Data for Surgical AI via Consistent Image-Label Generation}

\author{Niran Nataraj\corref{cor1}\fnref{label1}}
\ead{niran@g.ecc.u-tokyo.ac.jp}
\ead[url]{https://niran.dev}

\author{Maina Sogabe\corref{cor1}\fnref{label1}}
\ead{fsogabe-vet@g.ecc.u-tokyo.ac.jp}
\ead[url]{http://www.bmc.ipc.i.u-tokyo.ac.jp/index_e.html}

\author{Kenji Kawashima\fnref{label1}}
\ead{kenji\_kawashima@ipc.i.u-tokyo.ac.jp}
\ead[url]{http://www.bmc.ipc.i.u-tokyo.ac.jp/index_e.html}

\cortext[cor1]{Equal Contribution}
\affiliation[label1]{organization={The University of Tokyo, Graduate School of Information Science \& Technology},
            addressline={Bunkyo-ku},
            city={Tokyo},
            postcode={113-8656},
            country={Japan}}

\begin{abstract}

Deep learning for surgical image analysis is constrained by limited data diversity, ethical restrictions, high acquisition costs, and the scarcity of accurately annotated bleeding datasets, making reliable bleeding localization particularly challenging. To address these limitations, we propose orGAN, a modular framework that generates realistic, fully annotated synthetic surgical bleeding images together with recoverable point-level supervision. To overcome the ethical and cost barriers of real surgical data, the pipeline is trained on silicone-based mimicking organs that reproduce tissue properties, using an enhanced StyleGAN3 generator coupled with Relational Positional Learning to embed bleeding coordinates, a Surgical Label Detection Algorithm to recover coordinate and mask labels, and a fine-tuned LaMa inpainting network to restore clean image-label pairs. To compensate for scarce annotations, Bleeding Alert Map (BAM) localization models are trained on a blend of synthetic and mimicking-organ images. The enhanced StyleGAN3 attains an Inception Score of 2.24, and the inpainting network preserves realism at a mean SSIM of 0.98. The blended model outperforms both single-source baselines, reaching a Dice coefficient of 0.635 on paired bleeding frames. With a stronger Pix2PixHD backbone, Dice rises to 0.893 with a centroid error of 6.3 pixels. On real Hamlyn surgical video, the blended model achieves a frame-level accuracy of 0.858 and a bleeding-frame detection rate of 0.998 on Hamlyn 2, where all 500 frames were annotated as bleeding-positive. These results show that combining realistic synthetic images with mimicking-organ data is an effective, ethical, and economical strategy for developing surgical AI under limited annotated data.

\end{abstract}


\end{frontmatter}

\section{Introduction}\label{sec:intro}

Surgical artificial intelligence (AI) depends on large, well-annotated image datasets. Yet, for intraoperative bleeding, such labeled data remain scarce, costly, and ethically constrained to obtain.
Image synthesis can help enlarge these datasets, but existing generators rarely preserve the spatial labels that bleeding-localization models require, leaving a gap between realistic image synthesis and training-ready annotation.

Deep learning has revolutionized medical imaging by delivering substantial gains in segmentation accuracy, lesion detection, and diagnostic classification across radiology and endoscopy \cite{Litjens2017_MedIA, Liu2019_LancetDH, Hooper2023_NeurIPS}. However, translating these advances into surgical domains remains challenging due to the complex, dynamic environment of the operating room, where lighting variations, instrument occlusions, and rapid tissue deformations complicate robust model performance \cite{Bhatt2021_MMS, MaierHein2017_NatBME, Giannarou2013_TPAMI}. In particular, intraoperative bleeding, an event that can threaten patient safety if not detected promptly, poses a unique challenge: bleeding patterns vary unpredictably across patients, anatomical sites, and surgical techniques \cite{Schaefer2000_AJSurg, Gopinath2014_IJA}. Compounding these technical hurdles are severe data limitations, as acquiring and annotating high-fidelity surgical videos is resource-intensive, ethically constrained, and often restricted by patient privacy regulations \cite{Dhar2023_TTS}.

To mitigate clinical challenges and data limitations \cite{Schaefer2000_AJSurg, Gopinath2014_IJA}, our laboratory previously developed a multilayer silicone-based mimicking-organ system capable of reproducing realistic bleeding under endoscopic imaging conditions \cite{Sogabe2023_Array}. Endoscopic imagery commonly exhibits non-uniform illumination, low contrast, and color shift due to point-source lighting, which complicates both human interpretation and downstream learning \cite{HUANG2022106800}. By carefully layering silicone substrates and embedding colored fluid channels, we generated a dataset of annotated bleeding events that enabled the training of a Bleeding Alert Map (BAM) model with promising localization accuracy on \textit{ex vivo} and \textit{in vivo} test cases. Despite these successes, the manual fabrication process, which requires precise control of layer thickness, pigmentation, and channel geometry, demands several hours per sample, resulting in limited anatomical diversity and prohibitive scaling costs for larger datasets.

Addressing this bottleneck, we propose a structured data augmentation framework that seamlessly orchestrates generative modeling, relational positional encoding, automated label extraction, and inpainting into a unified pipeline for synthetic surgical image creation. By embedding anatomically plausible bleeding coordinates within a modified StyleGAN3 generator, extracting these coordinates via an automated detection algorithm, and applying advanced inpainting to remove residual artifacts, our approach yields high‐quality, artifact‐free images annotated with precise point‐coordinate labels. This scalable pipeline overcomes ethical and logistical barriers to data acquisition and enables training of localization models under severe data scarcity. Experimental validation demonstrates that blending orGAN-generated data with mimicking-organ data improves bleeding-localization performance over either source alone, highlighting the potential of our method for advancing surgical AI applications.

Motivated by the scarcity of labeled bleeding data, orGAN automatically produces paired (image, annotation) training data from a controlled physical phantom setup. Its novelty lies in combining Relational Positional Learning (RPL) marker embedding with Surgical Label Detection Algorithm (SLDA) extraction, which synchronizes labels \emph{through} the generation process itself rather than requiring pre-existing dense annotations. Evaluation on real Hamlyn surgical datasets confirms that the blended orGAN/mimicking-organ strategy improves BAM bleeding localization on real surgical video beyond models trained on either source alone.

\section{Current Scenario \& Related Work}
\label{sec:relworks}

The integration of AI into medical imaging has revolutionized diagnostics, treatment planning, and patient monitoring \cite{Litjens2017_MedIA, Esteva2017_Nature}. However, the application of AI, particularly deep learning, in medical imaging is constrained by the need for large, diverse, and accurately labeled datasets \cite{Tajbakhsh2020_MedIA}. In surgical imaging, these challenges are compounded by the invasive nature of data collection, ethical restrictions, and the inherent complexity of operative scenes \cite{Hashimoto2018_AnnSurg}.

\subsection{Challenges in Medical Image Data Acquisition and Labeling}

Medical imaging data acquisition faces numerous challenges, primarily due to strict privacy protections, regulatory constraints, and the necessity of specialized expert annotations \cite{Hosny2018_NatRevCancer}. Regulations such as the Health Insurance Portability and Accountability Act (HIPAA) and the General Data Protection Regulation (GDPR) place significant barriers on data sharing practices, thus complicating dataset compilation and dissemination \cite{Evans2018_GenMed}. Additionally, medical image annotation demands specialized expertise, rendering the annotation process both expensive and time-intensive \cite{Tommasi2008_PRL, Gu2019_TBME, Raedsch2023_NatMI}. In surgical contexts, acquiring data is further complicated by dynamic intraoperative variables including variations in patient anatomy, surgical technique, lighting conditions, smoke interference, and instrument presence, each of which introduces significant variability \cite{MaierHein2017_NatBME}. Furthermore, obtaining precise ground truth annotations (e.g., exact bleeding locations) in real-time surgical conditions is challenging, resulting in datasets that are typically small, imbalanced, and unrepresentative of the entire scope of surgical complexity \cite{Twinanda2017_TMI, Johnson2019_JBigData}.

\subsection{Use of Physical Phantoms and Mimicking Organs}

To mitigate data scarcity and overcome ethical concerns, researchers have developed physical phantoms or ``mimicking organs,'' which replicate human tissue properties for imaging studies and surgical training applications \cite{Tabaczynski2018_PhD, Sogabe2023_Array}. These models are often fabricated using materials such as silicone, hydrogels, or 3D-printed polymers to simulate the mechanical, optical, and acoustic properties of human tissues. For instance, \citet{Sogabe2023_Array} developed multilayer silicone-based mimicking organs to perform controlled bleeding simulations, generating images coupled with accurate ground-truth annotations. While mimicking organs reduce ethical barriers and facilitate experimental reproducibility, their production remains costly and labor-intensive, requiring detailed layering and coloring techniques to achieve realistic textures \cite{Lantada2016_Biofab, Sogabe2023_Array}. Furthermore, these models might not fully capture the anatomical variability and pathological complexity observed in real patients. Practical constraints limit the volume and diversity of image data generated, and the lack of spontaneous biological variation can also hinder the representativeness of resulting datasets.

Physics-based simulators solve the Navier-Stokes equations with smoothed particle hydrodynamics to model blood as a free-surface fluid, but realism versus real-time performance remains a core trade-off. A recent system introduces a stepless adaptive-particle scheme that varies particle mass and radius by position (fine at surface, coarse inside), adds repulsive forces to prevent volume distortion, and augments the pressure and tension terms to better capture non-Newtonian behavior, yielding smoother bleeding surfaces during suction and topological change at interactive rates \cite{SHI2024108322}. Beyond physical fidelity, data integrity and privacy have also emerged as crucial concerns in the creation of synthetic medical datasets. Bergen et al. demonstrated that adversaries could infer whether specific patient data contributed to the training of a generative adversarial network (GAN) via membership inference attacks on discriminator outputs, with discriminator-side areas under the curve (AUCs) approaching 0.99, while generator-only attacks hovered near chance. They proposed monitoring a privacy-utility curve during training and halting at the elbow to balance realism with privacy protection \cite{BERGEN2024107910}. Such findings emphasize that even ethically sourced synthetic data demand governance safeguards to prevent re-identification or leakage \cite{Drechsler2011_Synth}.

\subsection{Generative Adversarial Networks (GANs) in Medical Image Synthesis}

GANs, introduced by \citet{Goodfellow2014_GAN}, have demonstrated effectiveness in generating synthetic yet realistic images through adversarial learning between generator and discriminator networks. Within medical imaging, GANs have been utilized extensively to expand datasets, rectify class imbalances, and synthesize images representing rare medical conditions \cite{Yi2019_MedIA}. For example, GAN-generated liver lesion images have significantly enhanced the performance of classification models \cite{FridAdar2018_Neurocomputing}. GANs have also produced realistic synthetic brain magnetic resonance (MR) images \cite{Han2018_ISBI}, and synthetic MR images have improved tumor segmentation \cite{Shin2018_SASHIMI}. Lightweight adversarial generators tailored to small medical datasets can improve both image fidelity and downstream task performance. LF-GAN couples skip-layer channel sampling for robust gradient transfer with adaptive efficient group attention in the discriminator and a radiomics-inspired gray-level co-occurrence matrix loss focused on liver parenchyma and tumor texture, reporting a Fr\'echet Inception Distance (FID), peak signal-to-noise ratio, and structural similarity index (SSIM) of 75.73, 22.77, and 0.74 on the MTM-HCC dataset and gains in classification AUC and specificity under class imbalance \cite{ZHENG2024108252}.
Despite these successes, GAN-based methods applied to medical imaging must ensure not only visual realism but also anatomical correctness and clinical validity, requirements that present considerable challenges \cite{Shin2018_SASHIMI}. Moreover, GAN training is often prone to instability and mode collapse, complicating its consistent application and requiring careful balancing between generator and discriminator \cite{Wang2021_CSUR}.

\subsection{Advancements of StyleGAN and Its Role in Medical Imaging}

To address some traditional GAN limitations, the original StyleGAN~\cite{Karras2019_StyleGAN} and its subsequent iterations were developed. StyleGAN2 and StyleGAN3 refined this style-based architecture, providing enhanced fine-grained control over the synthesis process, significantly improving training stability and reducing visual artifacts \cite{Karras2020_StyleGAN2, Karras2021_NeurIPS}. In medical imaging, progressively grown GANs have synthesized high-resolution retinal and brain MR images \cite{Beers2018_PGGANMed}, and StyleGAN-based retinal image synthesis has improved retinal disease classification \cite{Kim2022_SciRep}. Despite these achievements, deploying StyleGAN specifically in surgical contexts remains challenging due to the inherent variability of surgical scenes, dynamic interactions between tissues and surgical instruments, and significant domain differences between surgical and traditional medical imaging scenarios \cite{GarciaPeraza2017_CARE, MaierHein2017_NatBME}.

\subsection{Synthetic Data Generation for Surgical Applications}

Generating synthetic surgical images represents a critical need given ethical constraints and the practical difficulties of obtaining extensive real surgical data \cite{Rueckert2024_ComputBiolMed, Nwoye2025_PRL}. Existing research primarily targets specific tasks such as instrument segmentation \cite{Shvets2018_ICMLA} and workflow analysis \cite{Mascagni2022_AnnSurg}, and GAN-based synthesis for these tasks frequently struggles to render realistic tissue textures and accurately depict instrument-tissue interactions \cite{Yoon2022_MICCAI}.

Several approaches generate surgical images whose labels remain valid after synthesis. Unpaired image-to-image translation converts rendered simulator scenes into realistic laparoscopic images while preserving the simulator's ground truth \cite{Pfeiffer2019_MICCAI}, and later work extends this idea to temporally consistent video \cite{Rivoir2021_ICCV}, to video with controllable illumination (CIIGAN) \cite{Chen2025_CIIGAN}, and to diffusion-based refinement of simulator output \cite{Martyniak2025_SimuScope}. Sharan et al.~applied the same principle to images of a physical surgical simulator, keeping suture landmarks consistent as the images were translated to an intraoperative appearance \cite{Sharan2022_JBHI}. In these methods, the label comes from a source in which it already exists, either a simulator that models the scene or a frame that has already been annotated. Diffusion-based generators now produce highly realistic surgical images from text \cite{Nwoye2025_PRL} and surgical video from a single frame and motion cues (SurgSora) \cite{Chen2025_SurgSora}, but they provide no automatically generated bleeding-source localization labels. Methods that generate an image together with its mask rely on manually annotated dense pixel masks, used either to learn the mask output \cite{Zhang2021_DatasetGAN, Thambawita2022_SinGANSeg} or to train a mask-conditioned generator \cite{Qadir2022_BSPC}. For intraoperative bleeding, these label sources are hard to obtain: physics-based simulation of blood still trades realism against speed \cite{SHI2024108322}, and dense bleeding masks are costly to annotate.

Bleeding detection itself has moved from handcrafted color features \cite{Okamoto2019_SIVP} to deep networks that detect bleeding in surgical video \cite{Jiang2021_ICGIP, Horita2024_SurgEndosc} or segment bleeding regions at the pixel level \cite{Sunakawa2024_SurgEndosc}. These detectors are trained on manually annotated clinical video, for example 34,117 annotated frames for real-time detection of active bleeding in laparoscopic colectomy \cite{Horita2024_SurgEndosc} and 2,203 pixel-annotated frames for bleeding recognition in laparoscopic hepatectomy \cite{Sunakawa2024_SurgEndosc}. Because collecting and labeling clinical data is costly and ethically constrained \cite{Hosny2018_NatRevCancer, Thambawita2022_SinGANSeg}, synthetic bleeding images that come with their own annotations offer a practical way to expand these training sets.

\subsection{Limitations and Gap Analysis of Current Synthetic Data Approaches}
\label{sec:gap_analysis}

Despite encouraging progress, current methods in synthetic data generation for surgical imaging exhibit several significant shortcomings. Firstly, most methods primarily focus on image generation without simultaneously generating corresponding ground-truth annotations, limiting their applicability for supervised AI tasks, particularly those requiring precise localization, such as bleeding detection \cite{Nwoye2025_PRL, Chen2025_SurgSora}. Secondly, current synthetic images often lack sufficient realism and diversity to capture the full complexity and variability inherent to real surgical environments, including different organ anatomies, pathological conditions, and intraoperative dynamics involving surgical tools \cite{Rivoir2021_ICCV, Chen2025_CIIGAN, Martyniak2025_SimuScope}. Another critical gap involves embedding precise positional information (e.g., bleeding locations) during the synthetic image generation process, a capability that existing methods provide only for labels that already exist \cite{Sharan2022_JBHI, Zhang2021_DatasetGAN}.

Additionally, standardized evaluation metrics designed specifically for the medical domain are notably lacking. Traditional image quality metrics fail to capture clinical relevance adequately, leading to insufficient validation of synthetic images' diagnostic value \cite{Borji2019_CVIU}. Finally, ethical considerations and potential biases introduced through the generation and use of synthetic data remain under-addressed, posing challenges for fair, effective clinical deployment of AI models.

Whereas related biomedical AI work, from feature-fusion segmentation and classification \cite{Nizamani2026_DeepFusion, Bhatti2025_FFUNet, Aamir2025_BrainTumor} to diagnostic reliability \cite{Khan2025_MagneticSensor}, typically relies on dense expert annotations and on overlap-based metrics such as Dice and intersection over union (IoU) \cite{TahaHanbury2015_BMCMI}, intraoperative bleeding localization rarely affords such labels, motivating a synthesis strategy that recovers point-level supervision while remaining evaluable with these same metrics.

These highlighted gaps collectively point to the urgent necessity for novel synthetic image generation approaches explicitly designed for surgical applications. Such approaches must simultaneously produce high-quality synthetic images with accurate, embedded ground-truth annotations, fully capture the intricate complexity of surgical scenes, and employ robust, domain-specific validation metrics. Addressing these critical limitations is paramount to developing AI models capable of robust performance across diverse real-world surgical settings.

A range of image-synthesis methods exists, but they differ from orGAN in the kind of supervision they assume. Paired and unpaired translation, semantic synthesis, and diffusion-based control all condition on labels that must already exist, such as segmentation maps, paired or unpaired image domains, or external control signals. orGAN instead generates the supervision it needs: point-level bleeding coordinates recovered during synthesis, rather than the dense segmentation maps these methods presuppose. Table~\ref{tab:litlandscape} summarizes this distinction across the principal method families in terms of the supervision each requires, whether it preserves localization labels, and its relevance to surgical bleeding synthesis. Among them, Pix2Pix, its VGG-perceptual variant, SPADE, and Pix2PixHD admit a like-for-like comparison on our data and are evaluated as BAM localization backbones under a common protocol (Section~\ref{sec:domain_metrics}). The remaining families require supervision that orGAN is designed not to need and are included for context.

\begin{table}[htbp]
\centering
\caption{Literature landscape for surgical image synthesis. ``Spatial label type'' distinguishes dense/segmentation supervision from sparse coordinate-point supervision; ``Label kept?'' marks whether localization labels survive generation (\checkmark) or not ($\times$); ``Target domain'' is the application the method was developed for.}
\label{tab:litlandscape}
\footnotesize
\setlength{\tabcolsep}{5pt}
\renewcommand{\arraystretch}{1.25}
\rowcolors{2}{gray!12}{white}
\resizebox{\textwidth}{!}{%
\begin{tabular}{@{}
  >{\raggedright\arraybackslash}p{1.55cm}
  >{\raggedright\arraybackslash}p{1.95cm}
  >{\raggedright\arraybackslash}p{1.7cm}
  >{\raggedright\arraybackslash}p{1.9cm}
  >{\centering\arraybackslash}p{0.85cm}
  >{\raggedright\arraybackslash}p{1.45cm}
  >{\raggedright\arraybackslash}p{2.95cm}@{}}
\toprule
\textbf{Method family} & \textbf{Representative work(s)} & \textbf{Supervision required} & \textbf{Spatial label type} & \textbf{Label kept?} & \textbf{Target domain} & \textbf{Position relative to orGAN} \\
\midrule

Paired image-to-image translation &
Pix2Pix, Pix2PixHD~\cite{Isola2017_Pix2Pix, Wang2018_Pix2PixHD} &
Paired source-target images &
Segmentation &
\checkmark &
Natural images &
Relies on labels that already exist, which orGAN can generate instead \\

Unpaired translation &
CycleGAN, CUT, MUNIT~\cite{Zhu2017_CycleGAN, Park2020_CUT, Huang2018_MUNIT} &
Unpaired image domains &
None (image-level) &
$\times$ &
Natural images &
Carries no point labels, while orGAN preserves them through generation \\

Semantic image synthesis &
SPADE, OASIS~\cite{Park2019_SPADE, Schonfeld2021_OASIS} &
Dense semantic maps &
Segmentation &
\checkmark &
Natural / street scenes &
Needs dense masks in advance, which orGAN does not require \\

Diffusion-based spatial control &
ControlNet~\cite{Zhang2023_ControlNet} &
Prompts and control maps &
Segmentation / control map &
\checkmark &
General images &
Uses external control signals, and its efficiency is left to future work \\

Surgical synthetic data &
\cite{Rueckert2024_ComputBiolMed, Mascagni2022_AnnSurg} &
Task-specific data or simulation &
Segmentation &
$\times$ &
Surgical &
Focuses on segmentation rather than recoverable bleeding points \\

\rowcolor{blue!12}
\textbf{orGAN (ours)} &
\textbf{SG3 + RPL + SLDA + LaMa} &
\textbf{Point-labeled source images} &
\textbf{Coordinate point} &
\textbf{\checkmark} &
\textbf{Surgical} &
\textbf{Generates new synthetic images together with corresponding coordinate and mask labels from point-labeled source data} \\

\bottomrule
\end{tabular}%
}
\end{table}

\begin{figure}
  \centering
  \includegraphics[width=1.0\linewidth]{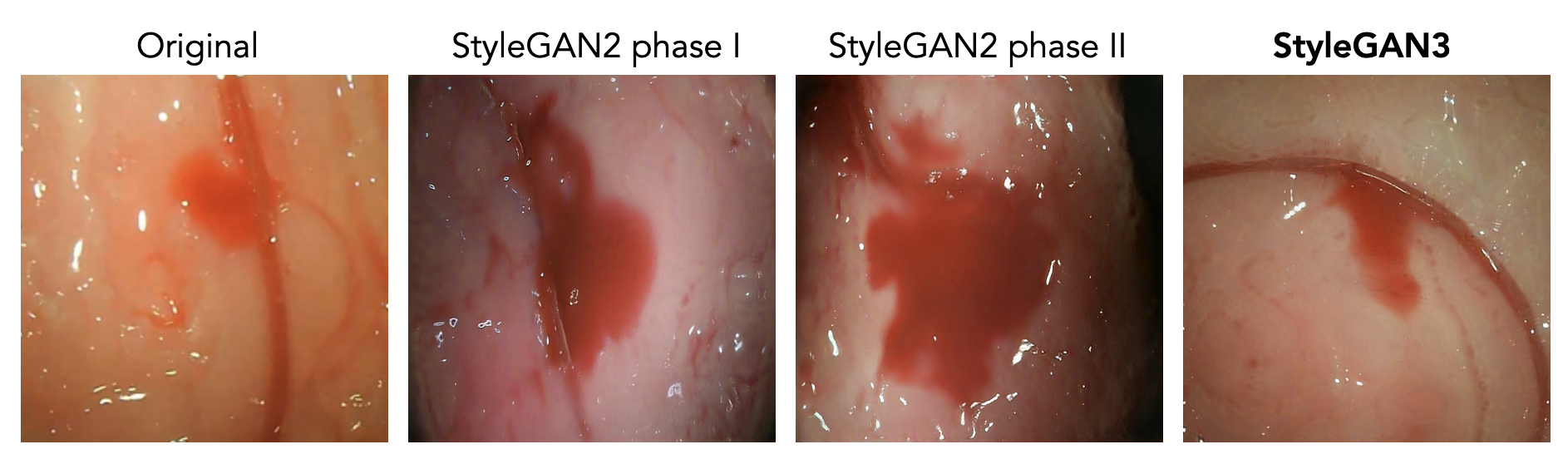}
  \caption{Generated sample images from the StyleGAN2 Phase I, StyleGAN2 Phase II, and StyleGAN3 models of the orGAN system, with an original mimicking-organ image for reference.}
  \label{fig1_app}
\end{figure}
\section{Contributions of This Research}

In response to the previously discussed challenges, we introduce \textbf{orGAN}, a multi-stage GAN-based framework specifically developed to generate synthetic surgical images annotated precisely for bleeding detection. Our core contributions include:

\textbf{1. Novel GAN Framework with Embedded Positional Labeling:}
We integrate RPL into a modified StyleGAN3 architecture (sample outputs in Figure~\ref{fig1_app}), embedding accurate bleeding coordinates into synthetic images. These annotations are reliably extracted using our proposed SLDA, creating ready-to-use annotated data for training localization models.

\textbf{2. Artifact-Free Image Generation through Optimized Inpainting:}
Leveraging advanced LaMa-based image inpainting, we effectively remove embedded labels post-extraction, ensuring realistic, artifact-free images suitable for diverse surgical applications beyond bleeding detection.

\textbf{3. Empirical Validation with Domain-Specific Metrics:}
Blending orGAN-generated data with mimicking-organ data yields the strongest bleeding-localization performance, reaching a Dice coefficient of 0.635 [95\% confidence interval (CI): 0.611, 0.659] on paired bleeding frames, a Hamlyn 1 frame-level accuracy of 0.858, and a Hamlyn 2 bleeding-frame detection rate of 0.998, outperforming both mimicking-organ-only and synthetic-only training.

\textbf{4. Ethical Scalability and Benchmarking of Synthetic Surgical Imaging:}
By substantially reducing reliance on real patient data and animal experimentation, orGAN provides an ethically sound and scalable solution to data scarcity. Moreover, the framework offers a reproducible reference point for synthetic surgical image generation, facilitating further advances in surgical AI applications.

\begin{figure}[t]
  \centering
  \includegraphics[width=1\linewidth]{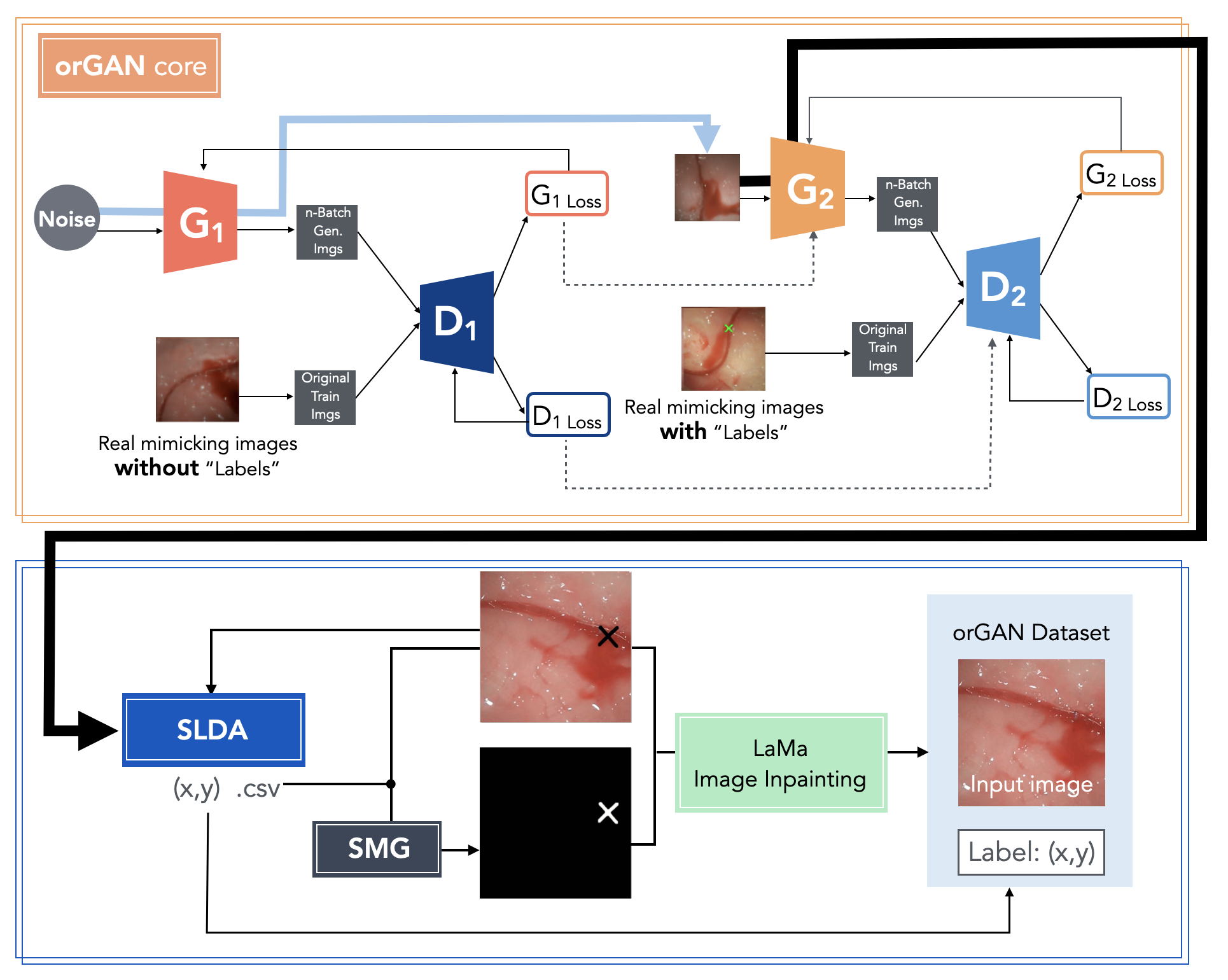}
  \caption{Overview of the proposed orGAN framework illustrating synthetic data generation, label embedding, extraction via SLDA, and subsequent inpainting to create realistic surgical training data.}
  \label{fig_overview}
\end{figure}

\section{Methodology}

We employ the mimicking-organ dataset \cite{Sogabe2023_Array} as our primary source for predicting bleeding locations in real intraoperative bleeding scenarios. Our proposed framework, referred to as orGAN, combines the strengths of GANs, RPL, SLDA, and advanced inpainting techniques to expand and refine the dataset.
At a high level, the pipeline executes seven stages in sequence: (i) acquisition of mimicking-organ bleeding images, (ii) StyleGAN3 synthesis, (iii) RPL embedding of recoverable point markers, (iv) automated SLDA coordinate extraction, (v) binary mask generation, (vi) LaMa inpainting to remove visible markers, yielding clean image-label pairs, and (vii) BAM training and evaluation on mimicking-organ and real Hamlyn surgical data (Figure~\ref{fig_overview} illustrates stages (ii) to (vi)).

\subsection{orGAN System}
The term ``orGAN'' reflects our core objective of generating organ-like synthetic images for medical applications, while drawing attention to the indispensable role of GANs in our pipeline. Traditional approaches often require labor-intensive physical setups or purely manual generation of synthetic data. By contrast, orGAN automates this process with a multi-stage GAN pipeline, effectively discovering and recreating the complex variability hidden within the underlying dataset \cite{Ciano2021_Math,Osman2022_BOE}.

Through the orGAN pipeline, we substantially enlarge the scope of synthetic data generation, producing organ-like images annotated with clinically relevant features. This approach not only conserves resources but also paves the way for robust, large-scale medical image datasets necessary for developing AI-based surgical support systems.

\subsection{Generating Synthetic Dataset: The Mimicking Organ Setup}

High-quality medical datasets are notoriously difficult to obtain due to various logistical, ethical, and privacy-related constraints. To address these challenges, prior studies have demonstrated that artificial organs made of layered silicone can closely replicate real tissues, including the realistic appearance of hemorrhagic events \cite{Sogabe2023_Array}. These ``mimicking organs'' are meticulously crafted, layer by layer, with accurate textures and colors, ensuring fidelity to true surgical scenes under an endoscope.

By leveraging these mimicking organs, we can systematically induce bleeding events under controlled conditions, facilitating the acquisition of images that mimic authentic intraoperative environments. As a result, we collect a diverse range of bleeding patterns without ethical hurdles or limitations on patient availability. This cost-effective strategy guarantees consistent image quality and standardized labeling, supporting robust AI training. The final evaluation of our system still targets real intraoperative bleeding scenarios, but the labeled training data primarily comes from these realistic, ethically sourced organ replicas.

The mimicking-organ system allows bleeding to be induced at selected locations under controlled conditions. The incision site was recorded as the bleeding-source coordinate in the image and assigned to the subsequent frames within the same video sequence. When geometric data augmentation (random cropping and horizontal flipping during BAM training) was applied, the same spatial transformation was also applied to the bleeding-source coordinate to preserve image-label correspondence.

\subsection{GAN over Mimic Data}

Although data from mimicking organs are remarkably close to real surgical scenes, their finite production can restrict both coverage and variability. GANs offer a powerful solution for augmenting and diversifying such datasets. In particular, structure-aware variants (SA-GAN) better maintain the geometry and arrangement of irregular anatomical features \cite{Emami2021_SAGAN}, thus helping the generator produce images consistent with real-world organ complexity.

\subsubsection{Training StyleGAN Models}

We performed initial experiments by training StyleGAN2 (SG2) and StyleGAN3 (SG3) on our mimicking-organ image set. StyleGAN2 is reputed for its ability to generate high-fidelity images while preserving a well-structured latent space \cite{Karras2020_StyleGAN2,Hermosilla2021_Access,Bermano2022_CGF}. Despite promising early trials, it occasionally showed inconsistencies that required vigilant monitoring, especially regarding temporal coherence.

To address these issues, we adopted a two-phase training approach, in which the first phase (PI) focused on establishing stable and consistent image generation, while the second phase (PII) aimed to improve robustness and cross-domain transfer.

Subsequent experiments using StyleGAN3 demonstrated enhanced temporal consistency and fewer aliasing artifacts, which are particularly important in medical imaging applications. Based on these observations, we selected the SG3 Phase II (SG3 PII) model for downstream components in our pipeline due to its visual stability and domain adaptability. Figure~\ref{fig1_app} shows sample images generated by these models.

\subsection{Inception Score }

The Inception Score (IS) \cite{Salimans2016_IS} evaluates both image quality and output diversity. It utilizes a pre-trained Inception network to classify generated images, assigning a high score when the per-image conditional class distribution is sharp (indicating realistic content) and when the overall marginal distribution is broad (reflecting diversity). Formally,
\[
\text{IS}= \exp\!\bigl(\mathbb{E}_{x\sim p_g}\,[D_{\mathrm{KL}}(p(y|x)\,\|\,p(y))]\bigr)
\]
where \( x \) denotes a sample drawn from the generator distribution \( p_g \), \( p(y \mid x) \) is the conditional class distribution for \( x \), \( p(y) \) is the marginal class distribution, and \( D_{\text{KL}} \) is the Kullback-Leibler divergence.

\subsection{Relational Positional Learning (RPL)}

\subsubsection{Concept of RPL}

RPL directly embeds label coordinates (e.g., bleeding points) into the training images, which the generator learns to reproduce, providing explicit spatial cues. This step is particularly valuable for medical applications that demand accurate localization of clinically relevant features such as hemorrhage sites.

This design also distinguishes the proposed framework from existing label-conditioned image synthesis approaches. Unlike label-conditioned image synthesis methods such as Pix2Pix, Pix2PixHD, and SPADE, which require pre-existing annotations as input, RPL embeds recoverable point markers directly into the training images, which the generator reproduces, and allows their automatic recovery through SLDA. Consequently, the proposed framework generates training-ready image-coordinate pairs without requiring dense annotations, making it suitable for bleeding-localization tasks where only sparse point supervision is available.\label{sec:novelty}

\subsubsection{Implementation of RPL}

To implement RPL, the supervision signal is embedded directly in image space rather than supplied as a separate channel or conditioning map. At every annotated bleeding coordinate $(x_i, y_i)$, a high-contrast marker glyph (an ``X'' rendered as two crossed line segments) is drawn onto the image prior to training. The generator \( G \) then learns to synthesize both realistic tissue and the embedded markers at their correct locations, while the discriminator \( D \) drives the markers to appear at consistent positions. SLDA later recovers them as coordinate labels. A high-contrast color keeps the markers separable from tissue tones during recovery, and black (RGB $(0,0,0)$) was selected for this purpose, as detailed in Section~\ref{sec:rpl_opt}.

Formally, let $I(x,y)$ denote the pixel intensity at location $(x,y)$, and let $\mathcal{X}(x_i,y_i)$ denote the set of pixels belonging to the ``X'' glyph centered at the $i$-th bleeding coordinate. The RPL-augmented training image $\tilde{I}$ is constructed as:
\[
  \tilde{I}(x,y) \;=\;
  \begin{cases}
    c_m & \text{if } (x,y)\in\displaystyle\bigcup_{i=1}^{n}\mathcal{X}(x_i,y_i),\\[4pt]
    I(x,y) & \text{otherwise,}
  \end{cases}
\]
where $n$ is the number of annotated bleeding sites and $c_m$ is the marker color (black, RGB $(0,0,0)$). The generator $G$ is trained on $\tilde{I}$, learning to reproduce both tissue appearance and the embedded glyphs at their correct spatial positions, from which SLDA recovers the marker coordinates.

\begin{figure}[t]
  \centering
  \includegraphics[width=0.7\linewidth]{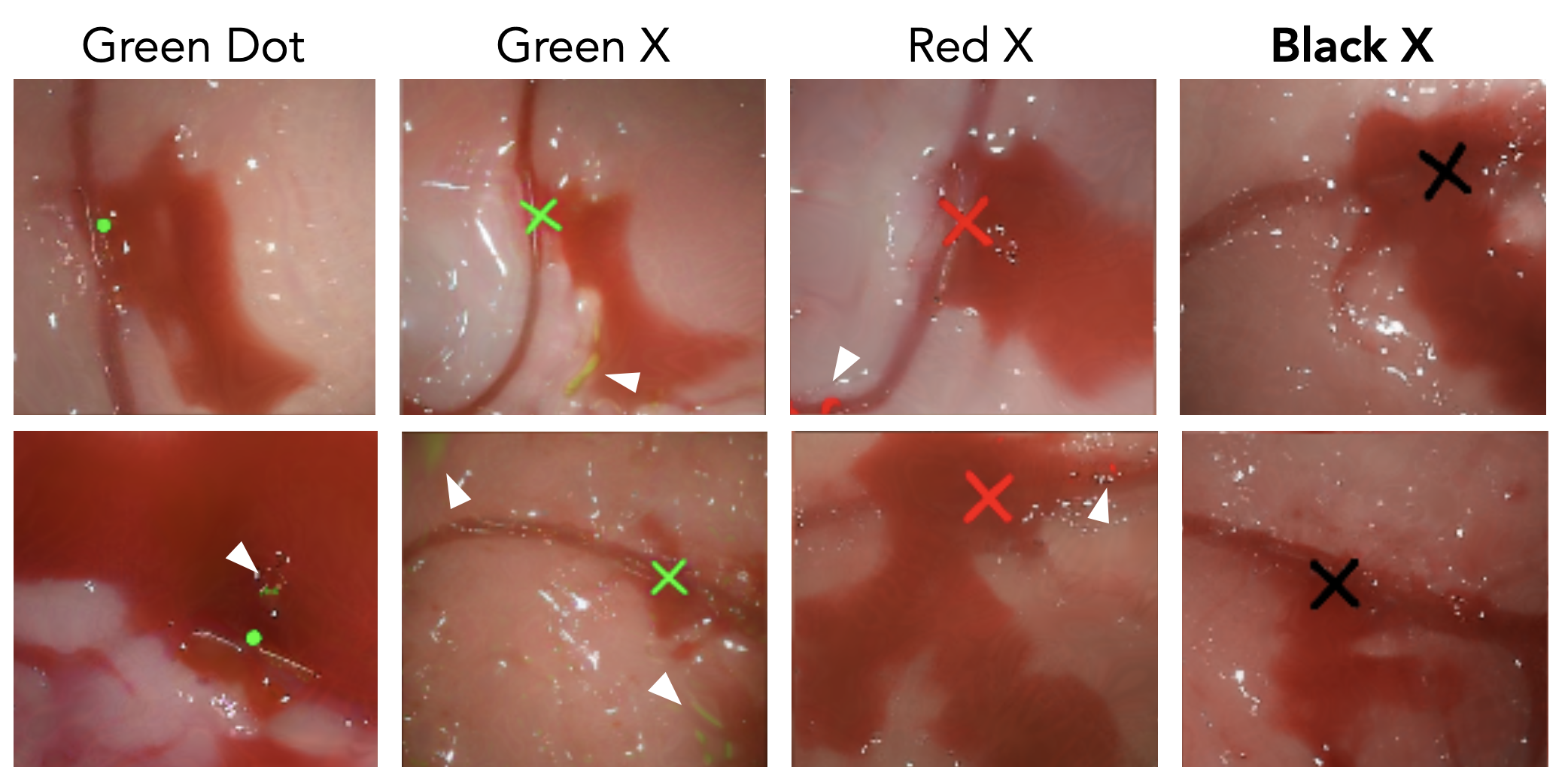}
  \caption{Labeled organ images generated by the orGAN system. White arrows highlight color spreading outside the designated label boundaries.}
  \label{fig:rpl_labels}
\end{figure}

\subsubsection{Training Objective}

The generator is trained with the standard StyleGAN3 objective: a non-saturating adversarial loss with an $R_1$ gradient penalty on the discriminator and path-length regularization on the generator. The discriminator drives the embedded markers to appear at consistent locations as part of matching the real marked distribution, and SLDA subsequently recovers them.

RPL additionally introduces a positional-consistency loss term, which is added to the adversarial loss to give the total training objective. Let \( \mathcal{L}_{\text{GAN1}} \) be the adversarial loss and \( \mathcal{L}_{\text{GAN2}} \) the positional-consistency loss introduced by RPL. The total loss becomes:

\[
\mathcal{L} = \mathcal{L}_{\text{GAN1}} + \lambda\,\mathcal{L}_{\text{GAN2}}
\]

where \( \lambda \) adjusts the importance of spatial alignment versus image realism.

\subsubsection{Transfer Learning and Phase Training}

To expedite convergence and stabilize training, we initialize the RPL network using weights from the best-performing SG3 PII model. This transfer-learning setup allows the network to focus on relational encoding, building on a generator already proficient in anatomical structure synthesis. This is analogous to the second learning phase of radial basis function networks \cite{Schwenker2001_RBF}, in which the hidden-layer parameters are fixed and only the output layer is trained.

The resulting model shows substantial improvement in spatial label fidelity, particularly for bleeding-localization tasks such as generating BAMs \cite{Sogabe2023_Array, osakabe2025comparative}.

\subsubsection{Challenges and Optimization}
\label{sec:rpl_opt}

Although RPL supports embedding multiple spatial markers, its performance degrades with dense labeling. We observed color leakage, particularly when using red markers, due to overlap with tissue tones. Switching to black (RGB $(0,0,0)$) significantly improved label contrast and reduced pixel interference. As shown in Figure~\ref{fig:rpl_labels}, high-contrast black markers showed less color leakage than green or red markers and therefore yielded better performance in SLDA and downstream analyses.

\subsection{Surgical Label Detection Algorithm (SLDA)}

\subsubsection{Purpose}

SLDA is designed to extract bleeding-point coordinates from GAN-generated images. These points are embedded during RPL using “X” markers, and SLDA provides an automated, accurate retrieval mechanism. The Segment Map Generator (SMG) is the part of SLDA that generates the binary marker mask used for inpainting (Section~\ref{sec:inpaint_proc}).

\subsubsection{Mathematical Formulation}

Let \( I \) be an input image, and \( \mathcal{I} \) the set of all images. SLDA is a mapping
\(
\mathcal{S} : \mathcal{I} \rightarrow \mathcal{P}(\mathbb{R}^2)
\)
where \( \mathcal{P}(\mathbb{R}^2) \) is the power set of 2D coordinate space, and the output is a set of bleeding label locations \( \mathcal{K} \subset \mathbb{R}^2 \).

\subsubsection{Algorithm Steps}

\begin{enumerate}
\item \textbf{Image Filtering:} Images with missing, poorly visible, or low-quality labels are discarded using automated heuristics. Visual inspection of 3{,}000 retained images confirmed that 2{,}984 (99.5\%) contained clearly recognizable markers.

\item \textbf{Color-space thresholding:} Convert \( I \) to the HSV color space and select the pixels lying within the marker's color range, producing a binary marker mask \( B \). For the black markers adopted here this isolates low-value pixels (\( V \leq 30 \)):
\[
B(x, y) =
\begin{cases}
  1 & \text{if } \mathrm{HSV}(I)(x, y) \in [\mathbf{c}_{\min},\, \mathbf{c}_{\max}], \\[3pt]
  0 & \text{otherwise.}
\end{cases}
\]

\item \textbf{Morphological Operations:} Apply morphological closing followed by opening with a \(5\times5\) kernel to consolidate the marker strokes and suppress speckle:
\(
B' = \mathrm{open}(\mathrm{close}(B)).
\)
\item \textbf{Contour Detection:} Identify contours in the processed binary mask:
\(
\mathcal{C} = \text{Contours}(B').
\)

\item \textbf{Shape Verification:} Each contour is simplified by polygon approximation (Douglas-Peucker, tolerance \(0.02\) of its perimeter) and accepted as an ``X'' marker when its approximation has between 5 and 19 vertices. A frame is retained only when the accepted-marker area is at least half of the detected mask area (confidence \(\geq 0.5\)).

\item \textbf{Centroid Calculation:} For each accepted contour \( c \in \mathcal{C} \) with \( N \) points \( (u_j, v_j) \), compute its centroid:
\[
(x_c, y_c) = \left( \frac{1}{N} \sum_{j=1}^{N} u_j,\; \frac{1}{N} \sum_{j=1}^{N} v_j \right)
\]

\item \textbf{Output:} Return the set of centroids as the label locations:
\(
\mathcal{K} = \{ (x_c, y_c) \mid c \in \mathcal{C} \text{ accepted in Step 5} \}.
\)
\end{enumerate}

\subsection{Image Inpainting}

\subsubsection{Purpose}

After SLDA extracts the bleeding-point coordinates from the “X” marks, we remove these visible artifacts from the images to obtain realistic, label-free synthetic images. This is essential for training segmentation or classification models that require clean visual input without overlaid annotations.

\subsubsection{LaMa Inpainting Architecture}

We employ LaMa~\cite{Suvorov2021_Lama}, a state-of-the-art image inpainting architecture that combines convolutional encoders with fast Fourier convolution layers. LaMa is known for its ability to seamlessly fill large masked regions while preserving structural and textural consistency. Tools built on the Segment Anything Model, such as Inpaint Anything~\cite{Yu2023_InpaintAnything}, further extend inpainting to prompt-guided selection. We fine-tune the pre-trained LaMa model on our mimicking-organ dataset to optimize label removal under surgical domain constraints.

\subsubsection{Inpainting Process}
\label{sec:inpaint_proc}

\paragraph{Mask Generation} First, a binary mask \( M \) is created over the label regions detected by SLDA:

\[
M(x, y) =
\begin{cases}
1 & \text{if } (x, y) \in \text{label region} \\
0 & \text{otherwise.}
\end{cases}
\]

\paragraph{Image Restoration} The masked image \( I \), along with the binary mask \( M \), is passed into the LaMa model to produce a clean inpainted image:
\(
I_{\text{clean}} = \text{LaMa}(I, M).
\)

This process eliminates embedded markers while preserving natural textures, so that the resulting images closely resemble the unannotated source images. These inpainted outputs are then suitable for training downstream medical AI models.

\subsection{Societal Benefits}
\label{sec:societal}

Beyond its technical contributions, orGAN carries broader societal value. By generating annotated synthetic datasets from controlled phantom setups, it reduces reliance on real patient data, limiting privacy risks and the ethical burden on clinical research, while reducing the manual expert-annotation bottleneck that constrains surgical AI. Because the data are generated from a controlled phantom-derived workflow, the pipeline supports reproducible benchmarking across groups without direct access to surgical data-collection infrastructure and may improve coverage of bleeding scenarios that are underrepresented in existing datasets. All datasets used in this study are de-identified, and no new patient or animal experiments were conducted.

\section{Experimental Results}
This section details the experimental evaluation of the proposed orGAN system, focusing on the quality of synthetic image generation, the marker configuration of the integrated RPL and SLDA pipeline, the effectiveness of the LaMa-based image inpainting, and the overall system's impact on the downstream task of BAM generation. Performance is further validated using actual surgical video datasets.

\subsection{Performance and Analysis of GAN Models}
The performance of the SG2 and SG3 models, trained on the mimicking-organ dataset across two training phases (PI and PII), was compared using IS and model size.

Among the models tested, SG2 PI and SG2 PII achieved IS values of 1.64 and 2.22, respectively, with model sizes of 7.8 GB each. SG3 PI and SG3 PII yielded IS values of 2.21 and 2.24, respectively, with model sizes of 8.5 GB each. IS provides a quantitative comparison of the generated images, computed with the base Inception model. Because the ImageNet-trained Inception-v3 network is out of domain for endoscopic scenes, IS should be interpreted only in relative terms, not against values reported for natural images \cite{Barratt2018_InceptionScore}. Based on these results, SG3 PII was selected for use in subsequent components of our pipeline due to its combination of the highest IS among the four base generators and stable generative performance, particularly in terms of visual consistency and temporal coherence, which are crucial for medical imaging applications.

\subsection{Evaluation of Marker Color Configurations in RPL-Based Label Extraction}

To determine the marker configuration that best supports label embedding and recovery, we evaluated the performance of the system under multiple marker color conditions. Specifically, we assessed how the choice of marker color affected both the image generation quality and training efficiency.

We tested four marker color configurations: green, red, green dot, and black. The IS values were as follows: green achieved 2.16, red 1.93, green dot 2.02, and black 2.27. In terms of training time, the black marker required only 41 hours, which was substantially less than green dot (312 hours) and green (175 hours), while being slower than the red marker (21 hours), a cost justified by its superior quality. The difference in training time reflects how easily the model learned the marker features of each color on a single GPU. Each configuration was monitored by IS, and training was stopped once the score had plateaued.

Based on these results, the black marker configuration was selected for use in subsequent experiments, as it achieved the highest generation quality of all configurations tested (IS 2.27), while remaining computationally tractable.

\subsection{Image Inpainting Results}

By using the fine-tuned LaMa model for image inpainting, we effectively removed the embedded labels from the images without introducing noticeable artifacts.

We measured the SSIM between the inpainted images and the original images (without labels). The average SSIM score was 0.98, indicating a high degree of similarity and confirming the effectiveness of the inpainting process.

\subsection{Evaluation of Properties within orGAN Images}
\begin{figure}[htbp]
  \centering
  \includegraphics[width=1.0\linewidth]{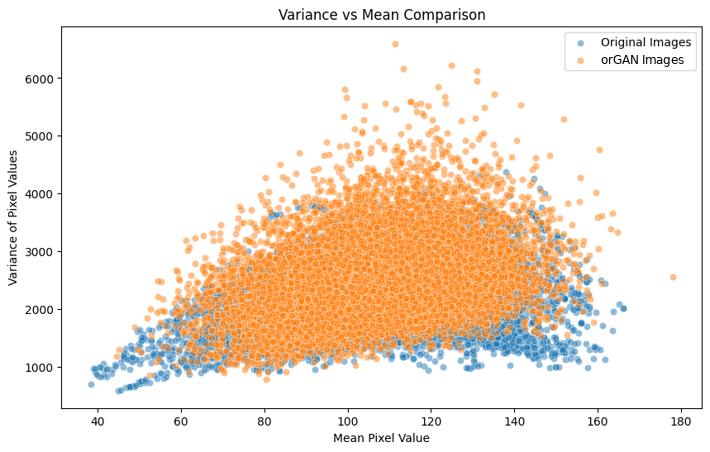}
  \caption{Variance versus mean pixel value for the original and orGAN-generated images.}
  \label{fig:var_mean}
\end{figure}

\begin{figure}[htbp]
  \centering
  \includegraphics[width=1.0\linewidth]{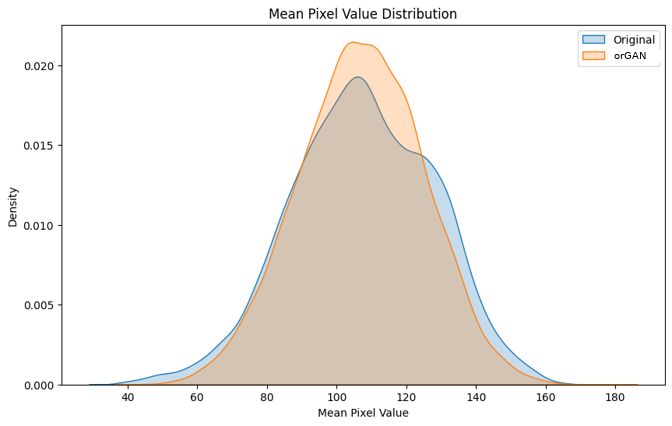}
  \caption{Kernel density estimate (KDE) of the mean pixel value for the original and orGAN-generated images.}
  \label{fig:kde}
\end{figure}

To evaluate the effectiveness of the orGAN model, we randomly selected 5,000 images from both the orGAN-generated and the original mimicking-organ datasets. For each image, key statistical properties were computed and subsequently visualized in Figures~\ref{fig:var_mean} and~\ref{fig:kde}.

Figure~\ref{fig:var_mean} displays the relationship between image brightness and contrast by plotting the variance of pixel intensities against the mean pixel intensity. The similar central tendencies of the orGAN-generated images and original images indicate that the overall brightness levels are well preserved, while the observed variance indicates that the synthetic images exhibit an acceptable degree of contrast variation, suggesting that orGAN is capable of generating novel images within the same domain.

Figure~\ref{fig:kde} presents a kernel density estimate of the mean pixel value distribution, providing a smooth representation of overall image brightness across both datasets. Both the original and orGAN-generated images show a unimodal distribution centered at a mean pixel value of approximately 110, indicating that the synthetic images preserve the overall brightness characteristics of the mimicking-organ data. The orGAN distribution is slightly more concentrated, suggesting that the GAN generates images with a more consistent brightness level, while the original dataset exhibits a marginally broader spread.

Collectively, these evaluations indicate that the orGAN-generated images adhere to the global intensity statistics of the original mimicking-organ images while adding new samples within the same domain.

\subsection{Bleeding Alert Map (BAM) Generation and Evaluation}

In this study, we evaluated the efficacy of the orGAN system in estimating bleeding locations during endoscopic surgeries, a task complicated by the difficulty of obtaining labeled datasets.

\subsubsection{BAM Generation Using Different Training Datasets}

\begin{figure*}[ht] \centering \includegraphics[width=1\linewidth]{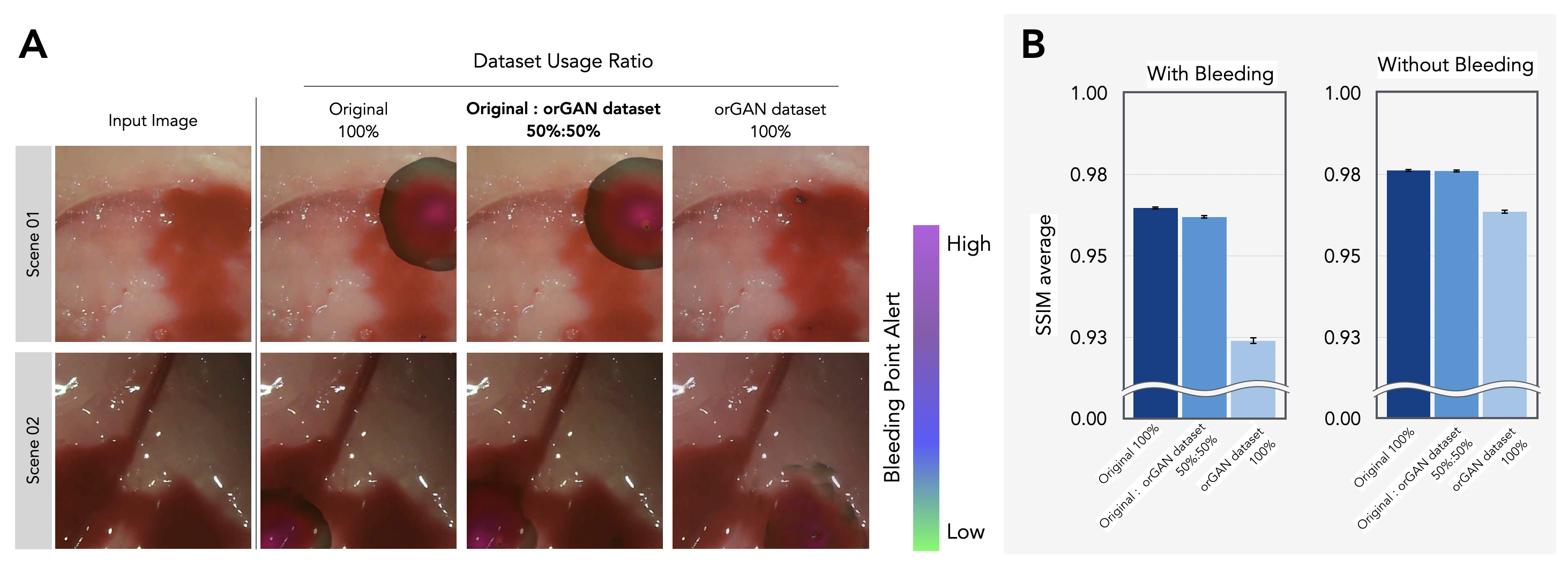} \caption{(A) Results of the BAM generated when training with varying ratios of datasets produced by orGAN (orGAN dataset) and primary datasets derived from mimicking organs (original). As input images, data from mimicking organs not used in the training process were utilized. (B) The average SSIM between the generated BAM and the ground truth is shown. The y-axis tick labels are rounded to two decimals (ticks at 0.925, 0.950, 0.975, and 1.000 above the axis break). The error bars represent the standard error.} \label{fig_bam_results} \end{figure*}

We trained BAM models using the Pix2Pix (paired image-to-image translation) architecture~\cite{Isola2017_Pix2Pix} on datasets with varying orGAN-to-original ratios. Figure~\ref{fig_bam_results}A contrasts the three representative configurations (the cross-configuration ablation in Section~\ref{sec:domain_metrics} also evaluates a 25:75 orGAN-to-original blend):

\begin{itemize}
\item \textbf{Original 100\%}: Only mimicking-organ images.
\item \textbf{orGAN 100\%}: Only images generated by orGAN.
\item \textbf{50:50 Blend}: An equal mix of the two datasets.
\end{itemize}

In the text, these configurations are also referred to as original-only, orGAN-only, and the 50:50 blend. In the Hamlyn result tables of Section~\ref{sec:realsurg}, they are denoted ORG100, GAN100, and ORG50:GAN50, and in the BAM result figures they are labeled by the percentage of each dataset used.

Each model was trained for up to 400 epochs, with the best checkpoint selected by validation Dice. Figure~\ref{fig_bam_results}A shows the BAM outputs for two randomly selected test images.

\subsubsection{Analysis of BAM Results}

\begin{figure}[htbp]
  \centering
  \includegraphics[width=1.0\linewidth]{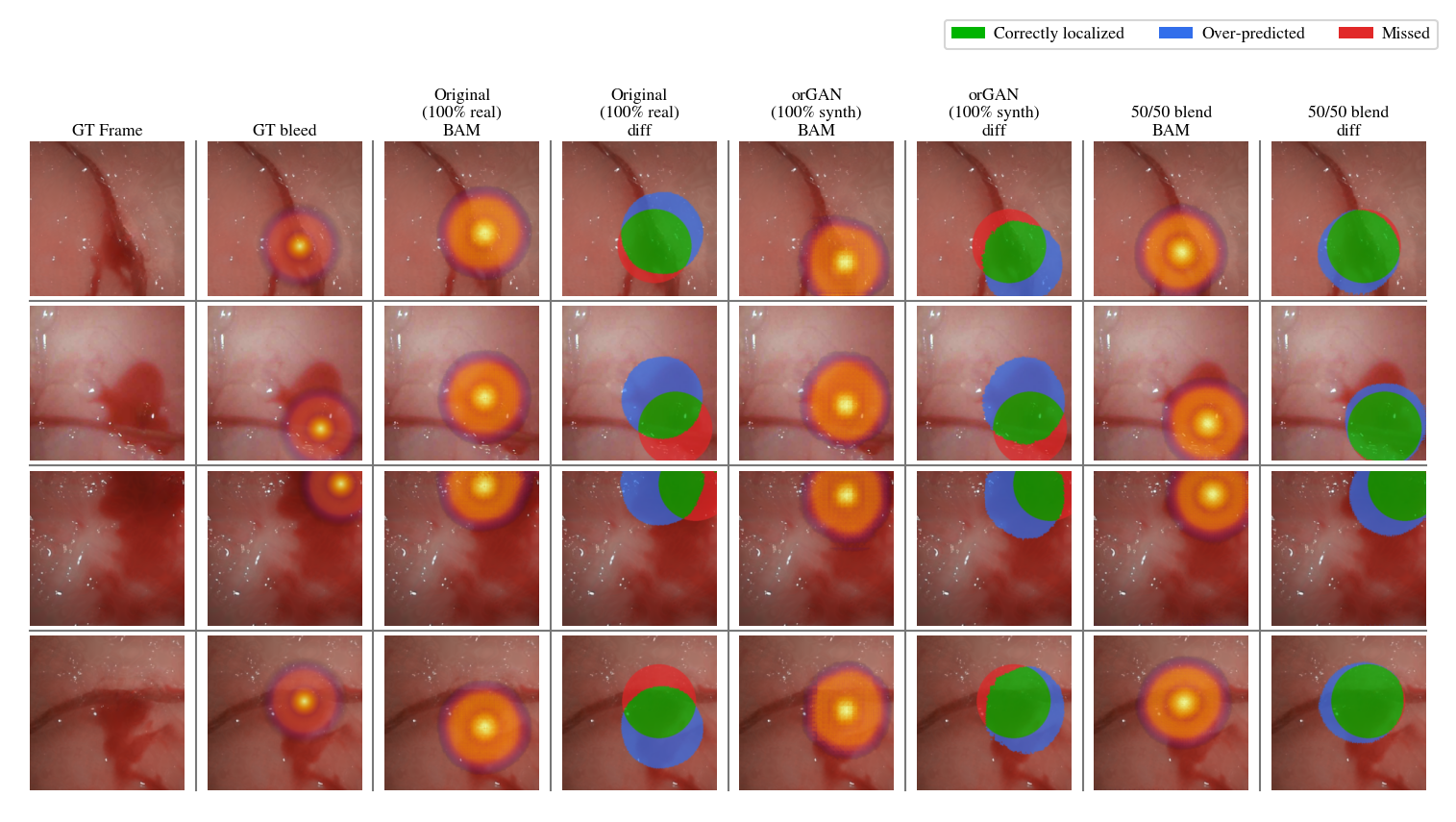}
  \caption{Qualitative bleeding localization on unseen mimicking-organ test frames. For each data configuration (Original mimicking-organ-only, orGAN synthetic-only, and the 50:50 blend) the predicted BAM heatmap (inferno colormap) and its overlap with the ground-truth region are shown. Overlap colors are defined in the key. Rows span a range of localization difficulty. See text for interpretation.}
  \label{fig:qual_panel}
\end{figure}

As shown in Figure~\ref{fig_bam_results}A, the model trained on the 50:50 blended dataset produced the most accurate BAMs, effectively identifying bleeding sources. The model trained exclusively on the original dataset performed adequately but less consistently, while the model trained solely on the orGAN dataset failed to produce precise BAMs.
Figure~\ref{fig:qual_panel} compares predicted bleeding heatmaps on unseen mimicking-organ test frames across the three data configurations. For each model, it also shows an overlap map against the ground truth, in which green marks correctly localized bleeding, blue marks over-prediction (predicted bleeding outside the ground-truth region), and red marks missed bleeding. The single-source models leave noticeably larger red and blue areas, whereas the 50:50 blend produces predominantly green overlap, indicating sharper and more spatially accurate localization, consistent with its higher Dice and IoU in Table~\ref{tab:ablation}.

\subsubsection{Quantitative Evaluation Using SSIM}

Figure~\ref{fig_bam_results}B presents the average SSIM scores for each model. The 50:50 blended model achieved average SSIM scores of 0.962 and 0.976 for images with and without bleeding, respectively. These values were comparable to those of the original-only model and higher than those of the orGAN-only model.

\subsubsection{Domain-Specific Localization Metrics}
\label{sec:domain_metrics}

To provide clinically meaningful evaluation beyond perceptual image quality metrics, we computed bleeding-localization performance on the paired test set using domain-specific metrics evaluated on BAM heatmaps thresholded at 0.10 with a 20-pixel minimum connected-component filter (the same protocol applied to the Hamlyn validation).
To make data composition the only factor that varies, all configurations in this ablation share an identical non-bleeding frame ratio (approximately 13\%) and identical validation and test splits. Only the source of the bleeding training frames (original, orGAN, or blended) varies across conditions.
All localization metrics operate on binary masks obtained by scalarizing each heatmap to its red channel, normalizing to $[0,1]$, thresholding at $\geq 0.10$, and discarding connected components smaller than 20 pixels (8-connectivity), applied identically to prediction and ground truth. Writing TP, FP, and FN for the true-positive, false-positive, and false-negative pixel counts, the overlap metrics are
\begin{align*}
\mathrm{Dice} &= \frac{2\,\mathrm{TP}}{2\,\mathrm{TP}+\mathrm{FP}+\mathrm{FN}}, &
\mathrm{IoU} &= \frac{\mathrm{TP}}{\mathrm{TP}+\mathrm{FP}+\mathrm{FN}}, \\[4pt]
\mathrm{Precision} &= \frac{\mathrm{TP}}{\mathrm{TP}+\mathrm{FP}}, &
\mathrm{Recall} &= \frac{\mathrm{TP}}{\mathrm{TP}+\mathrm{FN}}.
\end{align*}
Intuitively, Dice and IoU measure how well the predicted and true bleeding regions overlap, with IoU being the stricter of the two. Precision is the fraction of predicted bleeding pixels that are genuinely bleeding, and recall is the fraction of true bleeding pixels the model recovers. The centroid localization error is the Euclidean distance between the intensity-weighted centroids of the cleaned predicted and ground-truth bleeding regions, $e_c=\lVert \mathbf{c}_\mathrm{pred}-\mathbf{c}_\mathrm{gt}\rVert_2$, evaluated only on bleeding frames with a non-empty detection (complete misses are instead captured by recall). On non-bleeding frames, used only for the secondary false-alarm analysis, a correctly empty prediction is scored as perfect agreement by convention. All four metrics are computed for each frame and then averaged over the test set, so the values in Tables~\ref{tab:ablation} and~\ref{tab:arch} are per-frame means. Throughout, values shown in square brackets denote 95\% bootstrap confidence intervals.

For the blended (50:50) BAM model, the best-epoch checkpoint achieved a Dice coefficient of 0.635 [0.611, 0.659] on the bleeding-positive frames of the held-out test set (\mbox{$n=504$} of the 575 test frames), an IoU of 0.517 [0.494, 0.539], a precision of 0.567 [0.545, 0.589], a recall of 0.741 [0.714, 0.767], and a centroid localization error of 81.9 pixels [74.7, 89.1] on $256\times256$ frames with a non-empty detection.

Table~\ref{tab:ablation} reports the full cross-configuration ablation, spanning the two single-source corpora (mimicking-organ-only and synthetic-only) and the two blend ratios, each evaluated on the identical test frames with the same protocol, with Dice, IoU, and the remaining localization metrics reported alongside their 95\% bootstrap confidence intervals.

The 50:50 blend is the best configuration on every metric, and its Dice advantage is statistically significant. Paired Wilcoxon signed-rank tests (Holm-corrected) give $p = 2.8\times10^{-5}$ versus the 25:75 blend, $p = 1.5\times10^{-18}$ versus original-only, and $p = 1.8\times10^{-25}$ versus orGAN-only.
Critically, original-only and orGAN-only do \emph{not} differ significantly from each other ($p = 0.38$), which indicates that the performance gain arises from \emph{blending} mimicking-organ and synthetic data rather than from the choice of any single source corpus.
The 50:50 blend likewise leads on IoU, precision, recall, and centroid error, consistent with the Hamlyn surgical-video findings.
We deliberately report overlap- and localization-based metrics as primary rather than detection-only rates: because specular highlights and red tissue in endoscopic scenes closely mimic bleeding signal, a naive frame-level false-alarm rate on the paired mimicking-organ test set approaches 1.0 for \emph{all} four dataset configurations with the Pix2Pix backbone, and the models rarely abstain on non-bleeding frames. The clinically relevant question is whether the predicted bleeding region coincides with the true one, which Dice, IoU, and centroid error quantify directly.
Confidence intervals were computed via percentile bootstrap (10{,}000 resamples) on the per-frame metric distributions.

\begin{table}[htbp]
\centering
\caption{Cross-configuration ablation on the held-out test set ($n=504$ bleeding-positive frames). Values are means with 95\% bootstrap CIs (10{,}000 resamples) in brackets. The arrow marks the metric for which lower is better, and the best value in each column is in bold. Holm-corrected Wilcoxon tests against the 50:50 blend: $p = 2.8\times10^{-5}$ (vs.\ 25:75), $1.5\times10^{-18}$ (vs.\ original-only), $1.8\times10^{-25}$ (vs.\ orGAN-only). Original-only and orGAN-only do not differ significantly ($p = 0.38$). Protocol in Section~\ref{sec:domain_metrics}.}
\label{tab:ablation}
\setlength{\tabcolsep}{4pt}
\renewcommand{\arraystretch}{1.2}
\resizebox{\textwidth}{!}{%
\begin{tabular}{lccccc}
\toprule
\textbf{Configuration} & \textbf{Dice} & \textbf{IoU} & \textbf{Precision} & \textbf{Recall} & \textbf{Centroid err.}~$\downarrow$ \\
\midrule
orGAN-only (100\% synthetic)          & 0.469 [.444,.496] & 0.355 [.332,.377] & 0.412 [.388,.436] & 0.568 [.537,.598] & 142.3 [133.3,151.3] \\
Original-only (100\% mimicking organ) & 0.487 [.462,.513] & 0.372 [.350,.396] & 0.442 [.418,.467] & 0.566 [.538,.595] & 115.1 [107.4,123.0] \\
25\% orGAN + 75\% mimicking organ                & 0.576 [.551,.602] & 0.459 [.436,.483] & 0.522 [.498,.546] & 0.662 [.634,.690] & 92.7 [85.5,100.2] \\
\rowcolor{blue!12}
\textbf{50\% orGAN + 50\% mimicking organ}       & \textbf{0.635 [.611,.659]} & \textbf{0.517 [.494,.539]} & \textbf{0.567 [.545,.589]} & \textbf{0.741 [.714,.767]} & \textbf{81.9 [74.7,89.1]} \\
\bottomrule
\end{tabular}%
}
\end{table}

\begin{figure}[H]
  \centering
  \includegraphics[width=1.0\linewidth]{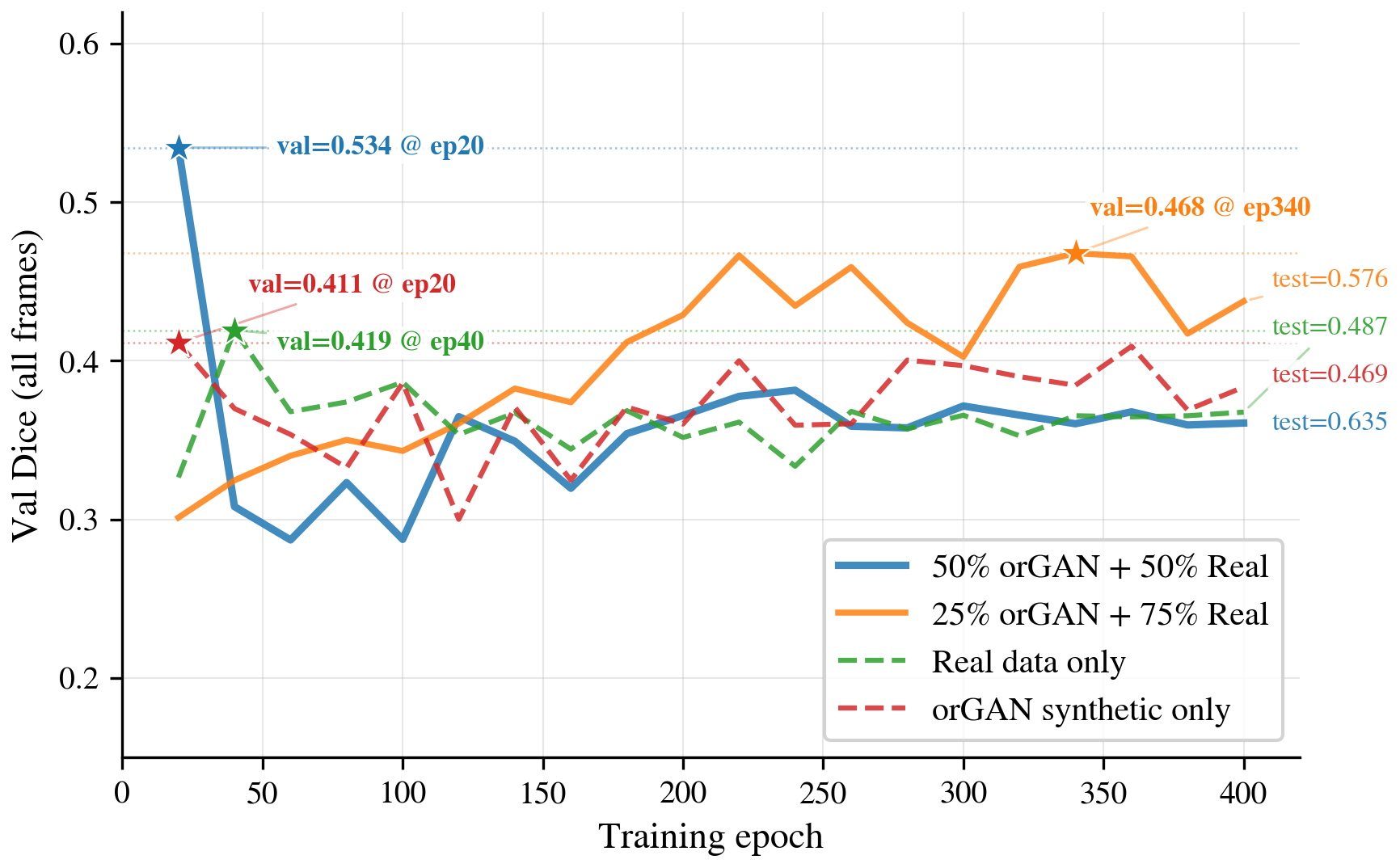}
  \caption{Validation-Dice trajectories across dataset configurations (Pix2Pix U-Net-256 backbone). Stars mark the best validation checkpoint used for test evaluation. Dashed lines indicate single-source baselines, and solid lines show blended training sets. See text for why the 50:50 blend is preferred.}
  \label{fig:epoch_traj}
\end{figure}

Figure~\ref{fig:epoch_traj} shows the validation-Dice trajectories across all four dataset configurations. The 50:50 blend reaches the highest validation peak, at epoch~20 (val-Dice~0.534), while the 25:75 blend improves more slowly and continues rising late in training. The original-only and orGAN-only baselines peak at lower values, confirming that neither source alone is sufficient.

Beyond the dataset ablation, we evaluated three architectural variants, each trained on the same 50:50 orGAN/mimicking-organ blend and evaluated on the identical held-out test set of 504 bleeding frames at a single fixed operating point (intensity threshold 0.10, 20-pixel minimum connected component). Each metric is reported as a mean with a 95\% bootstrap confidence interval. Centroid error is in pixels, and the frame-level false-alarm rate (FA) is the fraction of the 71 non-bleeding test frames on which a model predicts at least one blob. Table~\ref{tab:arch} summarizes the results. Adding a VGG perceptual loss to the Pix2Pix backbone raises test Dice from 0.635 to 0.881 [0.869, 0.893] and reduces centroid error from 81.9 to 18.7 pixels [15.5, 22.2], achieving high recall (0.956). However, the frame-level false-alarm rate remains 1.000, meaning the model triggers a positive prediction on every non-bleeding test frame. SPADE~\cite{Park2019_SPADE}, which re-injects the source image as spatially adaptive affine parameters at every decoder layer, achieves a test Dice of 0.845 [0.826, 0.864] and the lowest false-alarm rate of 0.070, correctly abstaining on 93\% of non-bleeding frames, roughly a 14$\times$ reduction in false-alarm rate relative to the Pix2Pix baseline (FA\,=\,1.000). Pix2PixHD~\cite{Wang2018_Pix2PixHD}, a multi-scale global/local generator with 182M parameters, achieves the highest test Dice of 0.893 [0.884, 0.901], the highest recall (0.984), and the lowest centroid error of 6.3 pixels [5.6, 7.1], a 13$\times$ improvement over the Pix2Pix baseline, while also substantially reducing false alarms (FA\,=\,0.169).

A further practical consideration weighs against SPADE despite its low false-alarm rate, and it concerns inference latency rather than accuracy. Because SPADE re-injects the source frame at \emph{every} decoder layer, it is markedly slower to run, producing a bleeding heatmap in over 300~ms per frame against roughly 100~ms for the feed-forward Pix2Pix backbone. This gap is decisive for intraoperative use, where the detector must keep pace with live surgical video. A surgeon can compensate for a $\approx$100~ms delay, whereas surgical performance degrades progressively beyond about 300~ms~\cite{Nankaku2022_Telesurgery, Xu2014_TelesurgeryLatency}. SPADE's added cost also does not yield better localization: its test Dice (0.845) is lower than Pix2PixHD's (0.893), and its only advantage is the lower false-alarm rate (0.070 versus 0.169). Training cost reinforces the same conclusion: SPADE's per-layer modulation forces a small training batch and a multi-day schedule to reach its best checkpoint, whereas the feed-forward Pix2Pix and Pix2PixHD models converge within a few hours. Pix2PixHD is therefore the recommended architecture. The Pix2Pix backbone is retained for the dataset-composition ablation to preserve continuity with the original BAM protocol and to isolate the effect of training-data composition. When stronger frame-level abstention is required, a calibrated confidence threshold or an explicit no-bleed head on the lighter backbone may offer a more economical route than adopting SPADE.

The behavior of the 25:75 blend warrants comment, because from about epoch~40 onward its validation curve lies mostly above the (early-peaking, then declining) 50:50 curve. We nonetheless adopt the 50:50 blend for three reasons. First, checkpoints are selected by best validation Dice rather than by final epoch, and the 50:50 peak (0.534) exceeds the 25:75 peak. This advantage carries over to the held-out test set, where the 50:50 blend attains a significantly higher Dice (0.635 versus 0.576, Table~\ref{tab:ablation}, Holm-corrected Wilcoxon $p = 2.8\times10^{-5}$) and leads on every other localization metric, so the higher validation curve of the 25:75 model does not translate into better held-out test performance. Second, the central objective of orGAN is to reduce dependence on manually produced mimicking-organ data: the 50:50 blend achieves the best localization while drawing only half of its training frames from mimicking-organ data, whereas the 25:75 blend needs three-quarters mimicking-organ data for lower accuracy, making the 50:50 blend preferable on both performance and data efficiency. Third, the far earlier convergence of the 50:50 blend (epoch~20 versus epoch~340) lowers training cost and reduces the risk of overfitting.

\begin{table}[htbp]
\centering
\caption{Cross-method architecture comparison on the 50:50 orGAN/mimicking-organ blend. Each metric is a mean with its 95\% bootstrap confidence interval (10{,}000 resamples) shown in brackets. Arrows mark metrics for which lower is better, and the best value in each column is in bold. The evaluation protocol and the false-alarm-rate definition are given in the text.}
\label{tab:arch}
\setlength{\tabcolsep}{4pt}
\renewcommand{\arraystretch}{1.2}
\resizebox{\textwidth}{!}{%
\begin{tabular}{lccccccc}
\toprule
\textbf{Architecture} & \textbf{Params} & \textbf{Dice} & \textbf{IoU} & \textbf{Prec.} & \textbf{Rec.} & \textbf{Cent.Err}~$\downarrow$ & \textbf{FA}~$\downarrow$ \\
\midrule
Pix2Pix~\cite{Isola2017_Pix2Pix} (50:50 blend)     & 54M & 0.635 [.611,.659] & 0.517 [.494,.539] & 0.567 [.545,.589] & 0.741 [.714,.767] & 81.9 [74.7,89.1]  & 1.000 \\
Pix2Pix + VGG Perceptual Loss                       & 54M & 0.881 [.869,.893] & 0.805 [.793,.817] & \textbf{0.821 [.809,.833]} & 0.956 [.944,.968] & 18.7 [15.5,22.2]  & 1.000 \\
SPADE~\cite{Park2019_SPADE}                         & 92M & 0.845 [.826,.864] & 0.770 [.752,.787] & 0.783 [.765,.799] & 0.929 [.907,.949] & 9.1 [8.2,10.2] & \textbf{0.070} \\
\rowcolor{blue!12}
\textbf{Pix2PixHD}~\cite{Wang2018_Pix2PixHD}       & 182M & \textbf{0.893 [.884,.901]} & \textbf{0.815 [.806,.823]} & 0.820 [.811,.827] & \textbf{0.984 [.975,.992]} & \textbf{6.3 [5.6,7.1]} & 0.169 \\
\bottomrule
\end{tabular}%
}
\end{table}

\subsection{Validation of Dataset Performance Using Actual Surgical Videos}\label{sec:realsurg}

To evaluate the efficacy of BAM in detecting bleeding in real surgical scenarios, we employed subsets of two publicly available datasets.

\subsubsection{Surgical Scene Datasets}

The \textbf{Hamlyn 1 dataset}~\cite{Giannarou2013_TPAMI} consists of dissection recordings of porcine diaphragms. These videos were captured at a resolution of $640\times480$ pixels and recorded at 30 frames per second. For our evaluation, we selected 500 consecutive frames with minimal smoke interference.

The \textbf{Hamlyn 2 dataset}~\cite{Stoyanov2005_MICCAI} comprises recordings of totally endoscopic coronary artery bypass graft (TECAB) procedures performed on human hearts using robotic-assisted endoscopy. In this dataset, the videos have a resolution of $348\times284$ pixels, and a 500-frame segment containing visible bleeding was selected for examination.

The two datasets differ from the mimicking-organ training data in species (porcine and human), organ (diaphragm and heart), and surgical procedure (diaphragm dissection and robotic-assisted TECAB). Evaluating on them therefore tests whether a model trained on mimicking-organ data transfers across domains rather than being confined to a single setting.

The clinical dataset was annotated and subsequently reviewed by a single surgical expert, whereas the mimicking-organ dataset was assigned coordinates systematically based on the known incision location. The visible bleeding locations in the clinical dataset were used as ground truth to evaluate model performance. When generating BAM, connected components of at least 20 pixels were retained. This 20-pixel threshold was selected as an approximate visual-support threshold. A 20-pixel blob corresponds to a region roughly 6 to 10 mm across in these videos, roughly the size of a small bleed that a surgeon can act on. A true positive was defined as the presence of BAM at locations where bleeding was apparent. A true negative was defined as the absence of BAM in images where no visible bleeding was observed, even if bleeding points were covered by instruments such as forceps.

\begin{figure}[htbp]
  \centering
  \includegraphics[width=0.72\linewidth]{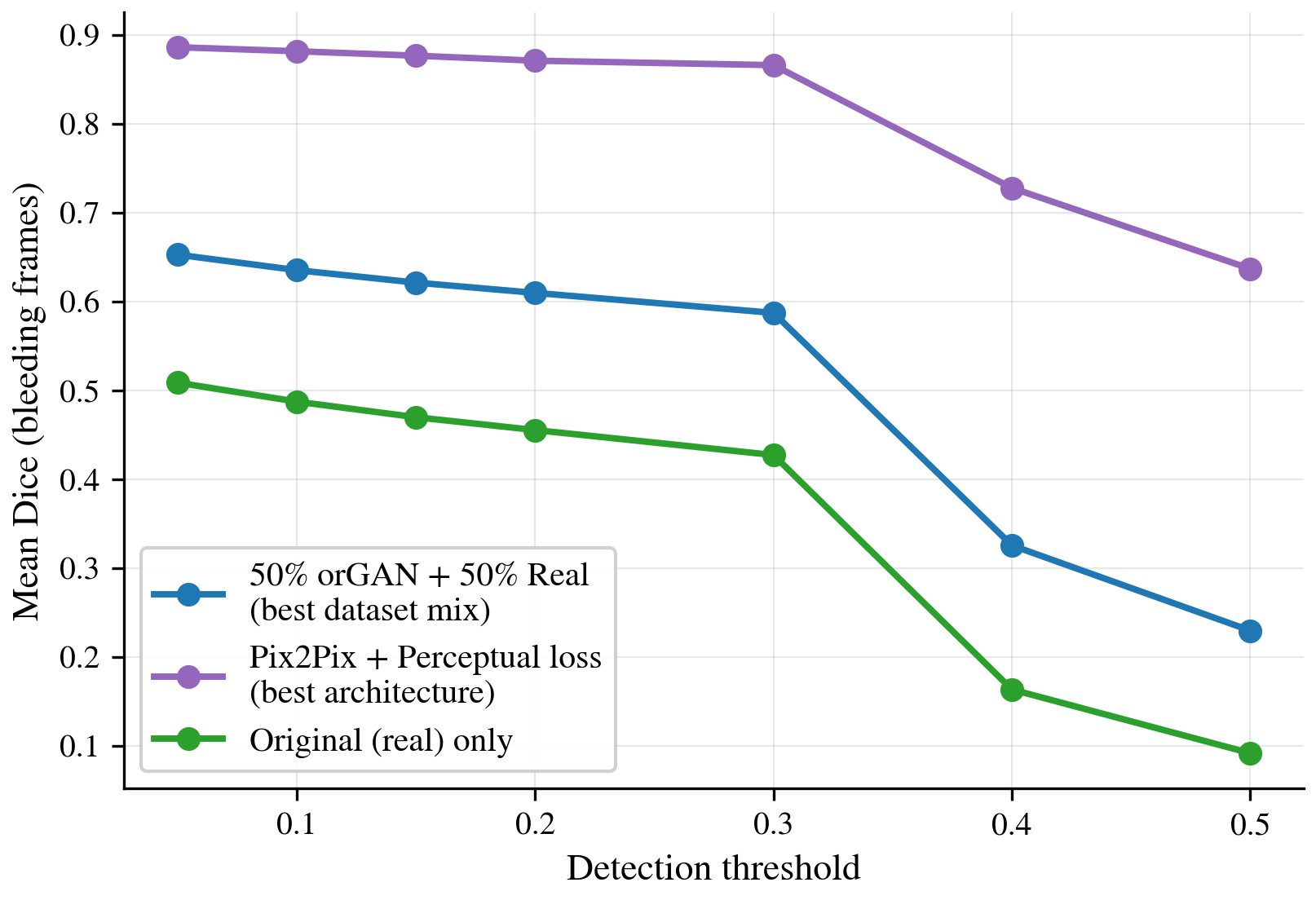}
  \caption{Detection-threshold sensitivity of Dice on the paired test set for the 50:50 blend (the perceptual-loss variant and the original-only model are shown for context). Dice is stable across low-to-moderate thresholds and falls only at aggressive ones, so the 0.10 operating point lies on a plateau.}
  \label{fig:thresh_sweep}
\end{figure}

The 500-frame evaluation windows were selected to span a contiguous segment of each recording with minimal smoke, maximizing annotation consistency.
To check that the 0.10 detection threshold and the 20-pixel blob criterion are not arbitrary, we swept both on the paired mimicking-organ test set (Figure~\ref{fig:thresh_sweep} shows the threshold sweep). The Dice coefficient is \emph{exactly} invariant to the minimum-blob size over 10-40~pixels, because at this threshold each predicted and ground-truth bleeding region is a single large connected component and the size filter removes nothing. The 20-pixel choice therefore has no effect on the reported Dice values. Dice also changes by only about 0.03 across detection thresholds from 0.05 to 0.15 (from 0.653 to 0.621) and declines sharply only beyond 0.3, so the 0.10 operating point sits on a gentle plateau rather than at a sensitive peak.

\begin{figure}[t] \centering \includegraphics[width=1\linewidth]{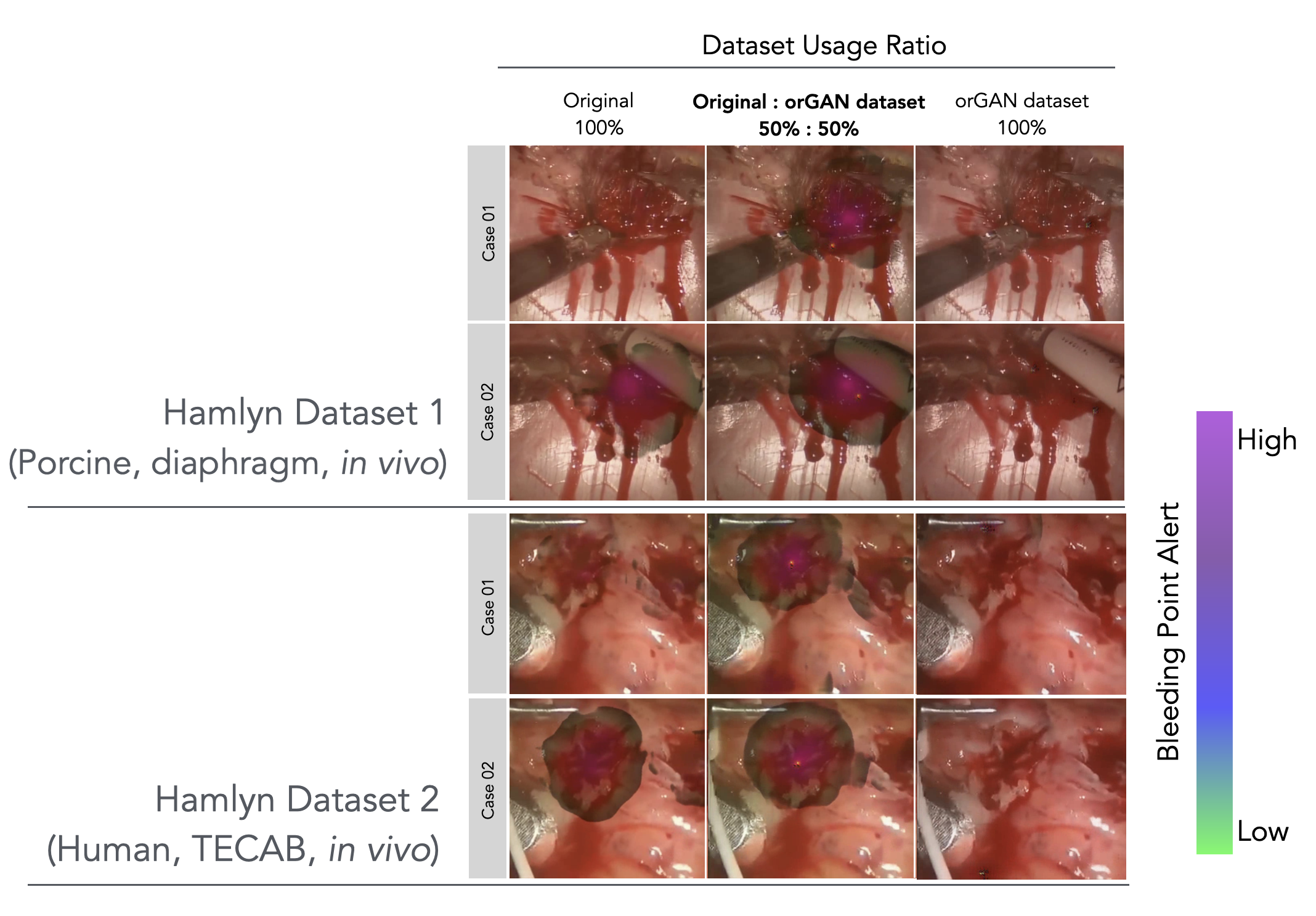} \caption{BAMs generated from actual surgical videos. The Hamlyn 1 dataset consists of dissection scenes of a porcine diaphragm, while the Hamlyn 2 dataset includes scenes from a robotic-assisted totally endoscopic coronary artery bypass graft (TECAB) surgery on a human. The columns show, from left to right, ``Original 100\%'' (only mimicking-organ images), ``Original : orGAN dataset 50\% : 50\%'' (an equal mix of the two datasets), and ``orGAN dataset 100\%'' (only images generated by orGAN). BAMs from the orGAN-only model are mostly mislocalized and are barely visible at this scale.} \label{fig_surgical_bam} \end{figure}

Figure~\ref{fig_surgical_bam} shows the BAMs generated from actual surgical videos.

\subsubsection{Analysis of Results}

The BAM generator trained exclusively on the orGAN dataset failed to produce precise BAMs in actual surgical scenes. When only the original dataset was used, BAM could be produced, albeit not for all frames, and its localization was less precise. In contrast, when orGAN-generated data were blended with the original mimicking-organ data, BAM was generated for most Hamlyn 1 bleeding frames and for almost all Hamlyn 2 frames.

For Hamlyn 1, frame-level accuracy is defined as the fraction of correctly classified frames, $(N_{\mathrm{TP}}+N_{\mathrm{TN}})/N$ with $N=500$, where $N_{\mathrm{TP}}$ is the number of frames with a correctly localized bleed and $N_{\mathrm{TN}}$ is the number of non-bleeding frames (including occluded frames) with a correctly empty prediction, as tallied in Table~\ref{tab:hamlyn001}. For Hamlyn 2, all 500 frames were annotated as bleeding-positive (Table~\ref{tab:hamlyn002}), so the 0.998 value is interpreted as a bleeding-frame detection rate rather than balanced detection accuracy or specificity. In the orGAN 100\% group, performance was extremely low (0.010 Hamlyn 1 frame-level accuracy and 0.156 Hamlyn 2 detection rate), and relying solely on the original data led to intermediate performance (0.586 Hamlyn 1 frame-level accuracy and 0.924 Hamlyn 2 detection rate). The 50:50 blend of orGAN and original data achieved the strongest Hamlyn performance, with a frame-level accuracy of 0.858 for Hamlyn 1 and a bleeding-frame detection rate of 0.998 for Hamlyn 2. This highlights the efficacy of blending orGAN-generated data with the original dataset. Across both real-surgery datasets, the blended model continued to localize bleeding despite the change in species, organ, and procedure, suggesting that the benefit of orGAN augmentation carries over to these settings rather than being tied to the mimicking-organ domain.

Tables~\ref{tab:hamlyn001} and~\ref{tab:hamlyn002} give the frame-level outcome counts of the three BAM training configurations on Hamlyn 1 and Hamlyn 2. Together, these results support the use of orGAN-generated data for improving bleeding-localization models under this evaluation protocol.

The publicly available datasets used for these evaluations were already de-identified and exempt from institutional review board review.

\begin{table}[h]
\centering
\caption{Performance for Hamlyn 1 (\protect porcine diaphragm dissection).}
\label{tab:hamlyn001}
\setlength{\tabcolsep}{2pt}
\renewcommand{\arraystretch}{1.1}
\small
\begin{tabular}{l >{\columncolor{blue!12}}c c c}
\toprule
Ground-truth status / BAM output & \textbf{ORG50:GAN50} & \textbf{ORG100} & \textbf{GAN100} \\
\midrule
Non-bleed (including occlusion) / No BAM   & 133 & 133 & 1   \\
Non-bleed (including occlusion) / BAM      & 0   & 0   & 132 \\
Bleed / Correct BAM  & 296 & 160 & 4   \\
Bleed / Incorrect BAM& 2   & 16  & 363 \\
Bleed / No BAM       & 69  & 191 & 0   \\
\bottomrule
\end{tabular}
\end{table}

\begin{table}[h]
\centering
\caption{Performance for Hamlyn 2 (human TECAB).}
\label{tab:hamlyn002}
\setlength{\tabcolsep}{5pt}
\renewcommand{\arraystretch}{1.1}
\small
\begin{tabular}{l >{\columncolor{blue!12}}c c c}
\toprule
Ground-truth status / BAM output & \textbf{ORG50:GAN50} & \textbf{ORG100} & \textbf{GAN100} \\
\midrule
Non-bleed / No BAM   & 0   & 0   & 0   \\
Non-bleed / BAM      & 0   & 0   & 0   \\
Bleed / Correct BAM  & 499 & 462 & 78  \\
Bleed / Incorrect BAM& 0   & 0   & 421 \\
Bleed / No BAM       & 1   & 38  & 1   \\
\bottomrule
\end{tabular}
\end{table}

\subsection{Reproducibility and Implementation Details}

Table~\ref{tab:repro} collects the implementation details. Training and evaluation used one NVIDIA RTX 5000 Ada GPU (32~GB), CUDA 11.6, and PyTorch 1.13.1. The BAM models were trained on the mimicking-organ data of~\cite{Sogabe2023_Array}, on orGAN-generated data, or on a blend of both. The 50:50 set combines orGAN and mimicking-organ frames in an approximately 1:1 blend, each frame stored side by side with its label map as a pair of $256\times256$ halves (12{,}599 training pairs), with a fixed split of 500 validation and 575 test frames reused across every model.

Upon acceptance, the source code of orGAN will be made publicly available at \url{https://github.com/kawashima-lab-ut/orGAN}. Links to the trained model weights used in this study are provided in the repository, and access to the mimicking-organ dataset is subject to dataset-sharing constraints.
\begin{table}[htbp]
\centering
\caption{Settings used to train and evaluate each component of orGAN.}
\label{tab:repro}
\renewcommand{\arraystretch}{1.2}
\resizebox{\textwidth}{!}{%
\begin{tabular}{@{}>{\raggedright\arraybackslash}p{2.7cm} >{\raggedright\arraybackslash}p{4.6cm} >{\raggedright\arraybackslash}p{5.6cm}@{}}
\toprule
\textbf{Component} & \textbf{Key parameters} & \textbf{Notes} \\
\midrule
StyleGAN3 (SG3~PII) & $256\times256$, Phase~II & IS 2.24. SG3 selected over SG2 \\
RPL marker embedding & Black marker, $\lambda=1$ & IS 2.27. Black gives the best quality-efficiency trade-off \\
SLDA / SMG & HSV value threshold $V\le 30$; closing then opening, $5\times5$ & 2{,}984/3{,}000 (99.5\%) retained images with recognizable markers \\
LaMa inpainting & Fine-tuned on mimicking-organ set & Mean SSIM 0.98 \\
BAM (Pix2Pix) & U-Net-256, batch 16, 400 epochs & Threshold 0.10, blob 20 px. Best epoch chosen via val-Dice \\
Training data (50:50 blend) & 12{,}599 train / 500 val / 575 test & $256\times256$, input frame to BAM direction \\
Hardware and software & RTX 5000 Ada, 32~GB, CUDA 11.6 & PyTorch 1.13.1, Python 3.9 \\
\bottomrule
\end{tabular}%
}
\end{table}

\section{Discussion: Limitations and Future Work}
\label{sec:discussion}

Despite these advances, several limitations warrant discussion. Broad cross-domain robustness of the orGAN system across diverse surgical environments remains to be established, as the current dataset may not encompass all real-world variations, such as differences in lighting conditions, organ textures, and surgical artifacts. Furthermore, the practical applicability is constrained by the specific type of synthetic organ used, necessitating the development of additional datasets to improve training diversity. The accuracy of labels generated by the RPL mechanism and SLDA may also be influenced by noise and annotation inaccuracies, thereby affecting the downstream AI performance.
To address these limitations, future work will focus on (i) expanding dataset diversity to better represent a wide range of surgical scenarios, (ii) improving computational efficiency to reduce overhead and increase training speed without compromising output quality, (iii) mitigating synthetic data bias by more closely approximating the complexity of real-world clinical data, and (iv) integrating alternative generator backbones within the present framework. Because label recovery and marker removal are performed post hoc by SLDA and inpainting, operations that are independent of the generator, a diffusion-based backbone could be substituted without altering the supervision-recovery pipeline, potentially enhancing synthesis fidelity while preserving automatic point-level annotation. A systematic comparison of generator backbones under a common protocol is left to future work.

Since generated data may ultimately be used in medical training or even clinical applications, ensuring their accuracy and reliability is critical. Errors in synthetic labels or features could lead to downstream misinterpretation or incorrect clinical decision-making.

In this study, we investigated the influence of marker color on image generation quality. To further enhance accuracy, future work will explore the impact of additional factors, such as the shape and size of markers, on SLDA and RPL performance. A systematic evaluation of these attributes is expected to contribute to the generation of more precise labeled data.

In addition to the conventional IS, we plan to incorporate alternative evaluation metrics for a more comprehensive assessment. For instance, the Conditional Fréchet Inception Distance (CFID) \cite{Soloveitchik2021_CFID} measures how well generated images conform to conditional input classes. The Kernel Inception Distance (KID) \cite{Binkowski2018_KID} offers a more stable alternative to FID \cite{Heusel2017_FID}, particularly for small datasets. Furthermore, the CLIP-based Maximum Mean Discrepancy (CMMD) \cite{Jayasumana2024_CMMD} leverages semantic embeddings and has demonstrated strong correlation with human visual judgment. The integration of these metrics will allow a more nuanced evaluation of generative quality.

Enhancements in RPL and SLDA will also be pursued to improve the spatial precision and robustness of label extraction. Moreover, establishing ethical guidelines and best practices for the development and dissemination of synthetic and mimicking-organ-based data will be essential to ensure transparency, reproducibility, and responsible use.

\section{Conclusion}

This study focused on addressing surgical imaging challenges, particularly the generation of synthetic data and its application for detecting bleeding. The orGAN framework, a multi-stage GAN pipeline, generates synthetic medical images with precisely annotated bleeding locations, while maintaining high fidelity. This approach tackles the scarcity of medical datasets, ethical concerns, and labeling inaccuracy. The orGAN system employs RPL and advanced image inpainting techniques to generate realistic surgical-scene images, thereby enhancing the training and performance of surgical AI models. Our testing revealed significant medical image synthesis capabilities of StyleGAN2 and StyleGAN3. The synthetic dataset generation using StyleGAN3, particularly in its second phase (SG3 PII), exhibited superior performance in terms of IS and stability. The selected generator achieved the highest IS among the four base generators (2.24), indicating the best relative fidelity and diversity among them. Beyond these scores, the generated images were of high visual quality.

 The BAM model trained with blended orGAN and mimicking-organ data achieved higher localization performance than the single-source configurations evaluated in this study. The combination of synthetic and mimicking-organ data yielded the most favorable results: the 50:50 orGAN/mimicking-organ blend achieved a Dice coefficient of 0.635 on paired bleeding frames, a Hamlyn 1 frame-level accuracy of 0.858, and a Hamlyn 2 bleeding-frame detection rate of 0.998 on a segment in which all 500 frames were annotated as \mbox{bleeding-positive}. Further model changes, namely adding a VGG perceptual loss or replacing the generator with SPADE or Pix2PixHD, substantially improved test Dice and localization precision, with Pix2PixHD reaching 0.893 Dice, a centroid error of only 6.3 pixels, and a false-alarm rate of 0.169, indicating that orGAN-augmented training data are compatible with stronger BAM architectures. This study demonstrates a reproducible approach to medical image synthesis using GANs and emphasizes the importance of ethical and scalable data generation. The orGAN system supports surgical assistance technology by providing a precise and ethically sound method for identifying and localizing bleeding in surgical video. This opens up opportunities for more sophisticated and reliable AI applications in the field of medicine.

\section*{Funding}
This work was supported by JSPS KAKENHI (Grant Numbers 25K21549 and 25H00717) and the JST ACT-X program (Grant Number JPMJAX23CC), Japan.

\section*{Declaration of competing interest}
Niran Nataraj and Maina Sogabe have no conflicts of interest or financial ties to disclose. Kenji Kawashima is the founder of Riverfield Inc.\ and holds shares in the company.

\section*{Data availability}
The Hamlyn surgical video datasets used for evaluation are publicly available from their original providers~\cite{Giannarou2013_TPAMI, Stoyanov2005_MICCAI}. The source code and the links to the trained model weights are provided in the repository at \url{https://github.com/kawashima-lab-ut/orGAN}. Access to the mimicking-organ dataset is subject to dataset-sharing constraints.

\section*{Ethics statement}
This study used publicly available surgical video datasets for evaluation purposes only. No clinical data were used for model training. The datasets contain no identifiable personal information and were made publicly available for research use by the original data providers. Therefore, institutional review board approval and informed consent were not required for this study.

\bibliography{sn-bibliography}

\subsubsection*{Declaration of generative AI and AI-assisted technologies in the manuscript preparation process}
\noindent
During the preparation of this work the author(s) used Generative Adversarial Networks in order to develop the model demonstrated in this paper and Grammarly to improve the sentence structure. After using this tool/service, the author(s) reviewed and edited the content as needed and take(s) full responsibility for the content of the published article.
\end{document}